\documentclass[aps,prb,reprint,twocolumn,superscriptaddress,showpacs,floatfix]{revtex4-1}

\usepackage{hyperref}
\usepackage{graphicx}
\usepackage{amsmath}
\usepackage{amssymb}
\usepackage{bbm}
\usepackage{color}
\usepackage{enumitem}

\definecolor{SI}{RGB}{180, 100, 30}

\begin{document}

\title{Self-energy embedding theory (SEET) for periodic systems}
\author{Alexander A. Rusakov}
\email{rusakov@umich.edu}
\affiliation{Department of Chemistry, University of Michigan, Ann Arbor, Michigan 48109, USA}
\author{Sergei Iskakov}
\email{siskakov@umich.edu}
\affiliation{Department of Physics, University of Michigan, Ann Arbor, Michigan 48109, USA}
\author{Lan Nguyen Tran}
\email{latran@umich.edu}
\affiliation{Department of Chemistry, University of Michigan, Ann Arbor, Michigan 48109, USA}
\affiliation{Department of Physics, University of Michigan, Ann Arbor, Michigan 48109, USA}
\affiliation{Ho Chi Minh City Institute of Physics, VAST, Ho Chi Minh City 70000, Vietnam}
\author{Dominika Zgid}
\email{zgid@umich.edu}
\affiliation{Department of Chemistry, University of Michigan, Ann Arbor, Michigan 48109, USA}
\affiliation{Department of Physics, University of Michigan, Ann Arbor, Michigan 48109, USA}
\affiliation{Center for Computational Quantum Physics, The Flatiron Institute, New York, NY 10010, USA}
\date{\today}
\begin{abstract}
We present an implementation of the self-energy embedding theory (SEET) for periodic systems and provide a fully self-consistent embedding solution for a simple realistic periodic problem - 1D crystalline hydrogen - that displays many of the features present in complex real materials. For this system, we observe a remarkable agreement between our finite temperature periodic implementation results and  well established and accurate zero temperature auxiliary quantum Monte Carlo data extrapolated to thermodynamic limit. We discuss differences and similarities with other Green's function embedding methods and provide the detailed algorithmic steps crucial for highly accurate and reproducible results. 
\end{abstract}
\maketitle
\section{Introduction}
For a solid described by a realistic Hamiltonian, a simultaneous illustration of weak and strong correlation is a difficult research problem and both density functional theory (DFT)~\cite{Kohn65} and ab-initio 
theories such as dynamical mean field theory (DMFT)~\cite{Georges96,Georges92,Georges04}, GW+DMFT~\cite{Biermann03,Biermann05,GW_review_werner2016}, density matrix embedding theory (DMET)~\cite{dmet_knizia12,dmet_knizia13,DMET_bootstrap_jcp_2016}, and more recently self-energy embedding theory (SEET)~\cite{Zgid15,Tran15b,Tran16,zgid_njp17,Tran_generalized_seet,Tran_GW_SEET,simons_benchmark2,Tran_useet} are being developed to address this challenge. 

SEET was initially introduced by us as a solver for the 2D Hubbard model~\cite{Zgid15} and later for molecular systems~\cite{Tran15b,Tran16,Tran_generalized_seet,Tran_GW_SEET,simons_benchmark2,Tran_useet}. We have demonstrated that SEET is capable of simultaneously addressing both weak and strong correlation and for molecular systems it displays some advantageous features such as a lack of intruder states and no necessity for higher order density matrices. In contrast, active space 
methods such as the complete active space perturbation theory second order (CASPT2)~\cite{roos1987complete,Kerstin:jpc/94/5483} require up to 4-body density matrices. The electronic energies obtained in SEET compare well with the ones yielded by standard quantum chemistry approaches~\cite{simons_benchmark2,Tran15b,Tran16}.  These features of SEET are beneficial when compared to typical quantum chemistry methods such as CASPT2 or the n-electron valence perturbation theory second order (NEVPT2)~\cite{Angeli1,Angeli2}. 

While the quest for ``an ideal theory framework'' can lead to multiple approaches, many corner stones are universally agreed upon. An ideal method ought to be systematically improvable and derivable from a theoretically sound framework. Additionally, any sources of errors should be controlled and removable provided that sufficient computational resources are available.

SEET is systematically improvable and  rigorously derivable since it can be expressed as a conserving functional that approximates the exact Luttinger-Ward functional~\cite{Luttinger60}, $\Phi_\text{LW}$. In SEET, to construct an approximate functional from all the orbitals present in the system, one chooses orbital groups that require more accurate or near exact description using a higher level method. The remaining orbitals or interactions between different orbital groups are evaluated by a lower level, less expensive method. Using the above reasoning, we have demonstrated that the general SEET functional~\cite{Tran_generalized_seet} can be written as
\begin{eqnarray}\label{eq:seet_mix_func} 
\Phi^\text{SEET}=&\Phi^\text{tot}_\text{weak}+\sum^{\binom{N}{K}}_{i}(\Phi^{A_i}_\text{strong}-\Phi^{A_i}_{weak})  \\ \nonumber
&\pm\sum_{k=K-1}^{k=1}\sum^{\binom{N}{k}}_{i}(\Phi^{B^k_i}_\text{strong}-\Phi^{B^k_i}_\text{weak}),
\end{eqnarray}
where the contributions with $\pm$ signs are used to account correctly for the possible double counting. 
$\Phi^\text{tot}_\text{weak}$ stands for an approximation to the exact $\Phi_\text{LW}$ functional  of the entire system evaluated using a weak coupling  method. $\Phi^{A_i}_\text{strong}$ denotes a functional obtained for an orbital group $A_i$ calculated using a higher level method able to describe strong correlations. $N$ stands for the total number of orbitals while $K$ is the number of orbitals within a single orbital group.
Note that this generalized form of SEET presents some similarities with the nested clusters scheme (NCS).~\cite{PhysRevB.97.125141}
This functional form of SEET can be further employed to write down self-energy expressions since the SEET self-energy is a functional derivate of the SEET functional with respect to a Green's function. 

Furthermore, if the orbital groups $A_i$ chosen are non-intersecting then the SEET functional from Eq.~\ref{eq:seet_mix_func} simplifies to
\begin{align}\label{eq:SeetPhi}
\Phi^\text{SEET} = \Phi^\text{tot}_{weak} + \sum_{i=1}^{M} \Big(\Phi_{strong}^{A_i}-\Phi_{weak}^{A_i}\Big).
\end{align}
The ``weakly correlated'' contributions to the functional can be evaluated by a low-order, here most frequently a perturbative method, such as self-consistent second order perturbation theory (GF2)\cite{Dahlen05,Zgid14,Rusakov16,Phillips15,Kananenka15,Kananenka16,Welden16,kananenka_hybrif_gf2,Iskakov_Chebychev_2018} or the GW method.\cite{Hedin65,Tran_GW_SEET}
The ``strongly correlated''  $\Phi^{A_i}_\text{strong}$ contributions to the SEET functional are evaluated for each of the orbital group/subspace $A_i$ by an accurate solver suitable for dealing with strongly correlated orbitals. Many choices of such solver are possible, ranging from quantum Monte Carlo methods~\cite{Gull11} to exact diagonalization type of solvers.
Here, we employ a solver based on truncated configuration interaction expansion~\cite{Zgid11,Zgid12}.

In this paper, we further extend the applicability of SEET to periodic systems and study its application to the simplest case, a 1D-periodic hydrogen solid.
We validate our prototype periodic implementation of SEET against accurate data obtained from auxiliary field quantum Monte Carlo (AFQMC)~\cite{PhysRevD.24.2278,PhysRevB.55.7464,PhysRevLett.90.136401,doi:10.1021/acs.jctc.7b00730}  to establish sources of errors and confirm that within the existing SEET approach quantitatively accurate solutions can be obtained.

This paper is organized as follows. In Sec.~\ref{sec:theory}, we introduce the theory and the necessary terminology for explaining the SEET algorithm for periodic systems. In Sec.~\ref{sec:algo}, we give a very detailed description of all the algorithmic steps. We discuss the orbital choice in Sec.~\ref{sec:orbitals} and the effects of choosing impurity orbitals in Sec.~\ref{sec:screening}. The computational scaling and exact limits of periodic SEET are discussed in Sec.~\ref{sec:limits_scaling}. Our results and a comparison with AFQMC is showcased in Sec.~\ref{sec:results}. We focus on a comparison of SEET with GW+EDMFT in Sec.~\ref{sec:dmft_comparison}. Finally, we form our conclusions and observations concerning future developments in Sec.~\ref{sec:conclusions}.
 
\section{Theory}\label{sec:theory}

In the real space, a Hamiltonian of a periodic system is defined as 
\begin{align}\label{eqn:ham}
\hat{H}&=\sum^{n_\text{cell}}_{g_i,g_j}\sum_{i,j}^{n_{orb}}t^{0,g_i - g_j}_{i \ j}a^{\dagger g_i}_{\ i}a^{g_j}_{j}+ \nonumber\\
 &\sum^{n_\text{cell}}_{g_i,g_j,g_k,g_l}\sum_{i,j,k,l}^{n_{orb}}v^{0, g_j-g_i,g_k-g_i,g_l-g_i} _{i\ j  \ k \ l}a^{\dagger g_i}_{\ i}a^{\dagger g_j}_{\ j}a^{g_l}_{\ l}a^{g_k}_{\ k},
\end{align}
where the real-space one-body and two-body translationally invariant integrals are given by $t^{0, g_i-g_j}_{i \ j}$  and $v^{0, g_j-g_i,g_k-g_i,g_l-g_i} _{i\ j  \ k \ l}$, respectively. The subscripts of $t^{0, g_i-g_j}_{i \ j}$  and $v^{0, g_j-g_i,g_k-g_i,g_l-g_i} _{i\ j  \ k \ l}$  refer to orbitals, here $i$ is an orbital index in a cell $0$, and $j$ is an orbital index in a cell $g_j$, etc. The superscripts refer to cells. 
The orbitals in this system may but do not need to be orthonormal and in general we will assume that they are non-orthogonal. The overlap matrix between orbitals is defined as $S^{0, g_i-g_j}_{i\ j}$. Note that in the later text we frequently denote $g_i-g_j=g$.

Here, for the sake of simplicity, we describe a grouping of orbitals in a traditional SEET scheme with non-intersecting impurities. 
For periodic systems, we assume that the N orbitals belonging to a unit cell can be separated into $M$ orbital subsets $A_i$,  each containing $N^A_i$ orbitals such that $N^A_i \ll N$. The the remaining $N^R$ orbitals are chosen to fulfill $N=\sum_{i=1}^M N_i^A+N^R$.
Note that when we write ``orbitals belonging to a cell", we actually do not mean that such orbitals are truly local to and contained in such a cell. Most of the time all the orbitals that we deal with are delocalized and spread out of the unit cell. In this context, the orbitals belonging to a cell  have a significant overlap with original atomic orbitals present in this cell.

If SEET is performed in the real space and the chosen groups of strongly correlated orbitals are non-intersecting then the SEET functional can be written as 
\begin{align}\label{eq:SeetPhi_periodic}
\Phi^\text{SEET} = \Phi^\text{tot}_\text{weak} + \sum_{i=1}^{M} \Big([\Phi_\text{strong}^{A \in \text{unit cell}}]_i-[\Phi_\text{weak}^{A \in \text{unit cell}}]_i\Big),
\end{align}
where all the chosen orbital groups $A_i$ belong to a unit cell in a crystal.  
Consequently, in the central real-space cell, the self-energy has the following form
 \begin{eqnarray}\label{eqn:sigma_seet}
 [\Sigma^\text{SEET}]^{0 0}=
 \begin{bmatrix}
    [\Sigma^\text{A}]_{1} & \Sigma^\text{int} & \dots &\dots &\dots\\
       \Sigma^\text{int}  &  [\Sigma^\text{A}]_{2} & \Sigma^\text{int} & \dots &\dots \\
     \dots & \dots &  \dots &  \dots &\dots\\
      \dots &  \dots &\Sigma^\text{int} & [\Sigma^\text{A}]_{M} &  \Sigma^\text{int} \\
  
   \dots &  \dots & \dots &    \Sigma^\text{int} & \Sigma^{R}
 \end{bmatrix},
 \end{eqnarray}
where different components of the central cell self-energy are defined as 
\begin{align}
[\Sigma^A]_{i}&=[\Sigma^\text{tot}_\text{weak}]^{0 0}+[([\Sigma^\text{A}_\text{strong}]_{i}-[\Sigma^\text{A}_\text{weak}]_i)]^{0 0},\label{eq:seet}\\
\Sigma^\text{R}&=[\Sigma^\text{R}_\text{weak}]^{0 0}, \label{eq:seetR}\\
\Sigma^\text{int}&=[\Sigma^\text{int}_\text{weak}]^{0 0} \label{eq:seetint}.
\end{align}
The real-space self-energy away from the central cell is defined as 
\begin{equation}
[\Sigma^\text{SEET}]^{0 g}=[\Sigma^\text{tot}_\text{weak}]^{0 g},
\end{equation}
where $g \ne 0$. Note, that here we use a notation introduced in Ref.~\citenum{zgid_njp17}.

Such a real-space definition results in the following k-space dependence of  the self-energy 
\begin{equation}
\Sigma^\text{SEET}(k)=[([\Sigma^{A}_\text{strong}]_{i}-[\Sigma^{A}_\text{weak}]_i)]^{00}+\Sigma_\text{weak}(k).
\end{equation}
The above definition stresses that the term $[([\Sigma^{A}_\text{strong}]_{i}-[\Sigma^{A}_\text{weak}]_i)]^{00}$ is k-independent, while all the other terms $\Sigma_\text{weak}(k)$ are k-dependent. 
Note that these relations hold for both frequency dependent and frequency independent components of the self-energy.

\section{General Algorithm}\label{sec:algo}
While the mathematical form of the SEET functional used for periodic systems given by Eq.~\ref{eq:SeetPhi_periodic} is simple, the algorithmic procedure involving the preparation of such a functional can be quite complicated. 
In this section, we list all the algorithmic steps necessary to perform a SEET calculation for a periodic system.

A periodic SEET calculation, analogous to molecular SEET, has two major parts: a calculation of the total system usually performed with a weakly correlated method and an iterative solution of an impurity problem performed with an accurate method capable of treating strongly correlated problems.

In the algorithm described below, we denote by {\bf LL} a low level loop and steps performed using a weakly correlated method. By {\bf HL} we denote a high level loop in which the impurity problem is solved.

\begin{description}

\item [LL0] Perform GF2 or GW in an iterative manner (if possible) until convergence on the whole periodic system.

\item [LL1] \label{FT_k_r} In k-space, evaluate  Green's function, $G(i\omega,k)$, self-energy $\Sigma(i\omega,k)$, Fock, $F(k)$, and  overlap,  $S(k)$ matrices. 
These are $n_k$ matrices of the size $n\times n$, where $n_k$ is the number of k-points, and $n$ the number of orbitals per unit cell. Note that due to the frequency dependence $G(i\omega,k)$ and $\Sigma(i\omega,k)$ are too large to be stored in memory and should be either evaluated on the fly or on a frequency grid with few points where the remaining points can be interpolated.~\cite{Kananenka16}
\item [LL2] Find the chemical potential $\mu$ to ensure that the number of electrons in each unit cell is proper.
\item  [LL3] \label{ortho_basis} Prepare matrices transforming to an orthonormal basis, $C^{SAO}(k)$, by using L\"owdin symmetric orthogonalization procedure. 
\item [LL3a--LL3d] Steps  are optional and only necessary when the basis of natural orbitals (NO) is used. Other bases such as molecular orbitals (MO), symmetrized atomic orbitals (SAO), etc. are also possible.
\item [LL3a] Calculate the one-body density matrix $\gamma^{\rm AO}(k)$.
\item  [LL3b] \label{dens_diag} Use the orthogonalizing basis $C^{SAO}(k)$ transformation to obtain the density matrix in an orthogonal basis
\begin{eqnarray}
\gamma^{\rm AO}(k) \to \gamma^{\rm SAO}(k).
\end{eqnarray}
\item [LL3c] Evaluate eigenvalues (natural occupation numbers) and eigenvectors (natural orbitals (NOs)) of the density matrix
\begin{eqnarray}
D(k)=(U^{DM}(k))^{\dagger} \gamma^{\rm SAO}(k)U^{DM}(k)
\end{eqnarray}
\item [LL3d] Prepare eigenvectors of the density matrix
\begin{equation}\label{dm_eigenvec}
\bar{U}(k)=C^{SAO}(k)U^{DM}(k).
\end{equation}
Note these eigenvectors transform from a non-orthogonal atomic orbital (AO) to the NO basis.
\item [LL4] Using  $C^{SAO}(k)$ or $\bar{U}(k)$  transform all the quantities listed below to the orthogonal basis (OR), which can be either SAO or NO  basis
\begin{eqnarray}
G^{\rm AO}(i\omega,k) \to G^{\rm OR}(i\omega,k),\\
\Sigma^{\rm AO}(i\omega,k) \to \Sigma^{\rm OR}(i\omega,k),\\
F^{\rm AO}(k) \to F^{\rm OR}(k).
\end{eqnarray}
\item [LL5]  Fourier transform all the quantities from the $k$- to the real-space. 
\begin{eqnarray}
G^{\rm OR}(i\omega,k)\to [G^{\rm OR}(i\omega)]^{0g},\\
\Sigma^{\rm OR}(i\omega,k)\to [\Sigma^{\rm OR}(i\omega)]^{0g},\\
F^{\rm OR}(k)\to [F^{\rm OR}]^{0g},\\
\gamma^{\rm OR}(k)\to [\gamma^{\rm OR}]^{0g},\\
\bar{U}(k)\to [\bar{U}]^{0g}.
\end{eqnarray}
Note that we want eigenvectors that have AO basis in the rows and an orthogonal basis (OR) in the columns. 
All these matrices in the real space have the dimension $n\times n\times n_{cell}$, where $n$ is the number of orbitals in the unit cell and $n_{cell}$ is the total number of cells used in the real space.
\\


\item [HL1] \label{embedding_loop1} If NO basis is employed, the real space density matrix   $[\gamma^\text{NO}]^{g_i g_j}$ has the following repeating motive 
 \begin{eqnarray}\label{eqn:dmr_non_diag}
 [\gamma^\text{NO}]^{g_i g_j}=
 \begin{bmatrix}
    X & 0 & 0 & \dots &\dots\\
       0  &  Y & 0 & \dots &\dots \\
     0 & 0 &  Z & \dots &\dots &\dots\\
     \dots & \dots &  \dots & X &0 &0&\dots\\
      \dots &  \dots &\dots & 0 &  Y &0 &\dots&\dots\\
  
   \dots &  \dots & \dots &   0 & 0 &Z &\dots &\dots
 \end{bmatrix},
 \end{eqnarray}
 where $X$ stands for doubly occupied orbitals, $Y$ are partially occupied orbitals with occupations significantly different than 0 or 2, and $Z$ are unoccupied orbitals with occupations very close to 0. The $X, Y, Z$ block is repeated in every unit cell. 
Note that the real space density matrix is diagonal in the $X, Y, Z$ blocks but non-diagonal everywhere else. The $[\gamma^\text{NO}]^{g_i g_j}$ is obtained as a result of a Fourier transform of the diagonal $\gamma^\text{NO}(k)$ matrix to the real space. For details see Sec.~\ref{sec:natural_wannier}.
 Choose $n_{act}$ orbitals of interest from the unit cell. For example, if NO basis is used, orbitals with partial occupations from the $Y$ block may be chosen. If SAO type of basis is used, criteria based on orbital spatial extent should be employed to choose best orbitals.
\item [HL2] Construct $n_{act}$ translationally invariant orbitals in the OR basis. The construction of these orbitals is described in Sec.~\ref{sec:natural_wannier}.
\item[HL3] \label{embedding_loop2}
Since all the quantities of interest are represented in an orthogonal basis (OR=NO or SAO), perform an embedding construction, where 
\begin{eqnarray}
[G(i\omega,k)]_{sub}=\bigg[\Big[(i\omega+\mu)\mathbf 1-F(k)-\Sigma_{tot}(i\omega,k)\Big]^{-1}\bigg]_{sub},
\end{eqnarray}
\begin{eqnarray}\label{k_space_sum}
[G(i\omega)]^{00}_{sub}&=\frac{1}{V_{BZ}}\sum_k\bigg[G(i\omega,k)\bigg]_{sub}e^{ikr=0},\\ 
\big{[}[G_0 (i\omega)]^{00}_{sub}\big{]}^{-1} &=[(i\omega+\mu)\mathbf 1-F]^{00}_{sub},
\end{eqnarray}
the hybridization of the chosen subset of orbitals with the rest of system is expressed as
\begin{eqnarray}\label{hybrid}
[\Delta(i\omega)]_{sub} &= \big{[}[G_0 (i\omega)]^{00}_{sub}\big{]}^{-1} -\big{[}[G(i\omega)]^{00}_{sub}\big{]}^{-1}-[\Sigma(i\omega)]^{00}_{sub}. 
\end{eqnarray}
Note that the size of the subset is $n_{act}$ which is the number of active/chosen orbitals. 
$[\Sigma(i\omega)]^{00}_{sub}$ is the subset of the total $[\Sigma^{OR}(i\omega)]^{00}$. This subset of the self-energy is constructed according to the Eq.~\ref{tot_se}. 

\item [HL4] Prepare an impurity model for the chosen orbitals. This requires two-body integrals in the OR basis for the impurity orbitals.
To obtain these integrals, denoted here as $v^{OR}_{ijkl\in sub}$, an integral transformation involving 2-body integrals and yielding impurity integrals has to be performed. Note that this transformation is  only necessary for the impurity integrals arising among $n_{act}$ orbitals.
In the real space, this operation scales as $O(n_{act}\times N^4)$ where $n_{act}$ is the number of impurity orbitals while $N=n_{cell}\times n$ is the total number of orbitals present in a periodic system. This operation can be further speed up by going to k-space and using density fitted integrals. 

\item [HL5]  Find the hybridization function from Eq.~\ref{hybrid} and evaluate $\epsilon_k$ and $V_{ij}$ that best fit the hybridization to the desired threshold according to the equation
\begin{equation}\label{hybrid_explicit}
\Delta_{ij}\approx\sum_{k}^{M}\frac{V_{ik}V_{kj}}{i\omega-\epsilon_k}.
\end{equation}
This step is only necessary when working in an explicit Hamiltonian formulation with a finite bath. The number of bath orbitals is denoted as $M$.

\item [HL6]  Run a solver capable of dealing with the non-diagonal hybridizations. In this paper, we employ either a truncated configuration interaction (CI)~\cite{Zgid12,Zgid11} or exact diagonalization (ED) called full configuration interaction (FCI) in the quantum chemistry community
to find $[\Sigma_\text{strong}(i\omega)]_{imp}$. The dimension of this self-energy matrix is $n_{act}\times n_{act} \times n_{\omega}$.

\item [HL7]  Evaluate the double counting correction $[\Sigma_\text{weak}(i\omega)]_{imp}$  using $v^{OR}_{ijkl\in sub}$ in an OR basis. 
This double counting correction is evaluated only for the impurity orbitals.

\item [HL8]  Remove the double counting correction and prepare the new self-energy in the following way 
\begin{eqnarray}\label{se+imp}
[\Sigma(i\omega)]^{00}_{imp}&=[\Sigma_\text{strong}(i\omega)]_{imp}-[\Sigma_\text{weak}(i\omega)]_{imp}.
\end{eqnarray}
Note that this total impurity self-energy $[\Sigma(i\omega)]^{00}_{imp}$ is in the OR basis.\\

\item [HL9] Transform the total self-energy $[\Sigma(i\omega)]^{00}_{imp}$  to the k-space 
\begin{equation}
[\Sigma(i\omega)]^{00}_{imp} \to [\Sigma(i\omega,k)]_{imp}.
\end{equation}
Note that if $[\Sigma(i\omega)]^{00}_{imp}$ is only obtained  in the central cell then $[\Sigma(i\omega,k)]_{imp}$ is the same for every k-point. 


\item [HL10]  In the k-space, construct the total subset self-energy as 
\begin{eqnarray}\label{tot_se}
[\Sigma(i\omega,k)]_{sub}&=[\Sigma(i\omega,k)]_{imp}+[\Sigma_\text{weak}(i\omega,k)]_{sub}\\\nonumber
&=[\Sigma_\text{strong}(i\omega,k)]_{imp}+\\ \nonumber
&+[\Sigma_\text{weak}(i\omega,k)]_{sub}-[\Sigma_\text{weak}(i\omega,k)]_{imp}.
\end{eqnarray}
This subset self-energy contains both the impurity self-energy $[\Sigma_\text{strong}(i\omega)]_{imp}$ which we will call an embedded contribution and the ``embedding'' self-energy contribution obtained by a lower level method, here $[\Sigma_\text{weak}(i\omega,k)]_{sub}-[\Sigma_\text{weak}(i\omega,k)]_{imp}$. 
For details about the self-energy structure, see Sec.~\ref{sec:screening}.

\item [HL11]  Build a new Green's function in the k-space 
\begin{equation}
G(i\omega,k)=[(i\omega+\mu)S(k)-F(k)-\Sigma(i\omega,k)]^{-1}.
\end{equation}
Here, for indeces $i,j$ belonging to the chosen/active orbitals the total self-energy $\Sigma(i\omega,k)$ is constructed according Eq.~\ref{tot_se}. For the remaining orbitals the self-energy  should be constructed as
\begin{equation}
[\Sigma(i\omega,k)]_{pq \notin sub}=[\Sigma_\text{weak}(iw,k)]_{pq \notin sub},
\end{equation}
\begin{equation}
[\Sigma(i\omega,k)]_{m \in sub \ n \notin sub}=[\Sigma_\text{weak}(iw,k)]_{m \in sub \ n \notin sub}.
\end{equation}

\item [HL12]  Find a new chemical potential $\mu$ to ensure a proper number of electrons in each of the cells. 
\item [HL13] 
Continue with the update of the hybridization and iterations starting from the step~{\bf HL3}. Check for the convergence of the impurity self-energy in each of the iterations. 


\item [LL6]  Transform the updated $G(i\omega,k)$ to the real space $[G(iw)]^{0g}$.  Using this Green's function perform a single iteration of GF2 according to the algorithm described in Ref.~\onlinecite{Rusakov16}. 


\item [LL7] Go to point~{\bf LL1}. 
\end{description}

The lower level embedding loop, if possible, should be carried out to self-consistency. In strongly correlated cases, it is usually important that the low level iterations are performed self-consistently since they lead to an adjustment of the weakly correlated orbitals and possible reevaluation of natural orbitals in the presence of strong correlations recovered from the impurity problem. Thus, in practice, while performing the lower level may be cumbersome, the potential improvement in the energy or properties may outweigh the complications.

\section{Orbitals for the impurity problem}\label{sec:orb_for_imp}
The solution of the impurity problem is supposed to improve on the initial solution coming from the weakly correlated method that was applied to the full periodic problem. Thus, it is very important that the orbital basis in which the impurity problem is expressed contains the most essential correlations present in the unit cell. 
This requirement may be expressed somewhat more formally. An ideal orbital basis is separating the full strongly correlated intractable problem into a weakly correlated problem for the whole solid and a strongly correlated problem involving few either localized or delocalized orbitals. In such an ideal orbital basis, the hybridization given by Eqs.~\ref{hybrid} and \ref{hybrid_explicit} is minimized, and the weakly and strongly correlated orbitals are almost decoupled. Consequently, solving the impurity problem in such a basis will result in recovering most  of the correlation contained in the unit cell. 

These requirements may result in different orbital choices depending on a specific physical situation. In a case when a physical system contains localized correlations, obviously, a good orbital choice is a localized orbital basis that will lead to small hybridizations. In other cases, when the correlations are non-local, e.g. equilibrium geometry where the orbitals have a significant overlap with each other and are delocalized, it may be necessary to employ orbitals that spread outside of the unit cell such as natural orbitals. We will discuss these choices in detail in Sec.~\ref{sec:results}.

Additionally, it is necessary that the orbital basis in which the impurity problem is expressed is orthogonal. In the non-orthogonal basis, the embedding construction involving Eqs.~\ref{hybrid}-\ref{k_space_sum} leads to a wrong behavior at high frequencies.
Moreover, since the embedding construction is done in the real space, it is necessary that the orbitals formed are translationally invariant in order not to break the periodic symmetry of the full problem. 
This means that the set of orbitals employed to form the impurity problem has to be the same in every cell. 

Consequently, Wannier orbitals are a natural choice since they form an orthogonal set and are the same for every cell of a periodic problem. However, we would like to stress that while maximally localized Wannier orbitals~\cite{RevModPhys.84.1419} are a common choice for periodic problems, in our case, we do not always employ localized Wannier orbitals, and the choice of a localized or delocalized basis is problem dependent. 
In this work, we employed two types of Wannier orbitals: regular and natural Wannier orbitals. 

\subsection{Wannier Orbitals}\label{sec:orbitals}

Here, we assume the general formulation of Wannier orbitals and we only demand that they form a set of orthogonal orbitals in the real space. 
We start by performing  the L\"owdin symmertic orthonormalization  of the atomic Bloch orbitals and obtain the transforming vectors in the k-space. These vectors diagonalize the overlap matrix in the k-space as follows:
\begin{equation}
[\delta(k)]_{ij}=\sum_{ij}[U(k)]_{i}^\dagger [S(k)]_{ij} [U(k)]_{j}.
\end{equation}
The vectors $[U(k)]_i$ are then Fourier transformed to the real space using
\begin{equation}
U^{0r}_{j}=1/V_{BZ}\sum_k [U(k)]_{j} e^{ikr}.
\end{equation}
The resulting real space vectors $U^{0r}_j$ are orthonormal and are periodic functions with respect to the real space cells. They are used as an orbital basis in which the impurity problem is defined.

\subsection{Natural Wannier Orbitals}\label{sec:natural_wannier}

Most commonly the independent particle Wannier orbitals are constructed as functions that preserve periodicity and diagonalize the Fock matrix coming from DFT or Hartree-Fock Fock calculations
\begin{equation}\label{eq:wannier_indep_ham}
[H(k)]_{ij}[u(k)]_j=[\epsilon(k)]_j [u(k)]_j.
\end{equation}
Consequently, the usual, independent-particle Wannier functions are formally defined as
\begin{equation}
w^{0r}_j=1/V_{BZ}\sum_k [u(k)]_j e^{ikr}.
\end{equation}
It is common that in many calculations these orbitals may be localized resulting in a maximally localized Wannier orbitals~\cite{RevModPhys.84.1419}.

Here, we employ a generalization of the independent particle Wannier orbitals to the case of interacting electrons since we found from our previous investigations involving SEET for molecular cases~\cite{Tran15b,Tran16,Tran_GW_SEET,Tran_generalized_seet,Tran_useet,simons_benchmark2} that such an orbital basis is excellent for impurity problems. The effective Hamiltonian from the eigenequation Eq.~\ref{eq:wannier_indep_ham} is replaced by a one-body density matrix $\gamma(k)$ in the following way:
\begin{equation}\label{dmk_definition1}
[\gamma(k)]_{ij} [U(k)]_j=[n(k)]_j [U(k)]_j.
\end{equation}
The above equation can be written alternatively as 
\begin{eqnarray}\label{dmk_definition2}
\gamma(k)=\sum_j [n(k)]_j [U(k)]_j [U^{*}(k)]_j,
\end{eqnarray}
where $[n(k)]_j$ are the natural occupation numbers in the k-space. In the real space, real space occupation numbers $[n^{00}]_j=1/V_{BZ}\sum_{k}[n(k)]_j$ preserve the periodicity and are repeated in every unit cell. Due to the N-representability condition in case of spatial orbitals the natural occupation numbers take the values $0\le  [n(k)]_j \le 2$.
Note that the eigenvectors from Eqs.~\ref{dmk_definition1}~and~\ref{dmk_definition2} are obtained according to Eq.~\ref{dm_eigenvec}. 
Subsequently, the correlated Wannier orbitals $W^{0r}_j$ are defined as
\begin{equation}\label{eq:natural_wannier_ft}
W^{0r}_j=1/V_{BZ}\sum_k [U(k)]_j e^{ikr}.
\end{equation}
This means that the correlated one-body density matrix in the real space is expressed as
\begin{equation}\label{eq:dm_r_wannier}
D_{ij}^{rr'}=\sum_{j,R,R'} n_j^{0(R-R')}[W^*]^{0(r-R)}_jW_j^{0(r'-R')},
\end{equation}
where $n_j^{0(R-R')}=1/V_{BZ}\sum_k [n(k)]_j e^{ik(R-R')}$. Note, that for independent particle Wannier only the term $n_j^{00}$ is present.

In SEET, the orbitals that are deemed important for the system of interest may be chosen based on the values of the natural occupation numbers $n_j^{0(R-R')}$ present in the real space. However, one should always remember that the selection should be both motivated by the natural occupation numbers as  well as physical/chemical characteristics of these orbitals. 

While the k-dependent one-body density matrix from Eq.~\ref{dmk_definition2} is diagonal in the k-dependent natural orbital basis, the real space density matrix from Eq.~\ref{eq:dm_r_wannier} in the natural Wannier orbital basis is non-diagonal and has the structure given by Eq.~\ref{eqn:dmr_non_diag}, where due to the Fourier transform, the off-diagonal blocks are present. The diagonal blocks of this matrix are diagonal and display a periodic pattern of occupation numbers $n_j^{00}$ that are present in every periodic cell. Thus, these natural Wannier orbital occupations can be used to help to determine the most correlated orbitals in a periodic cell. 
It is important to stress that this procedure keeps the growth of the number of correlated orbitals at bay since the number $n_{act}$ of chosen natural Wannier orbitals that are translationally invariant  from a periodic cell can be only between $0 \le n_{act} \le n$.

Note that this is not the case when CASSCF type calculations are done for the $\Gamma$-point in periodic problems. In such calculations, when natural orbitals are produced they are coming from multiple primitive cells present in the $\Gamma$-point calculation. Thus, the number of significantly correlated orbitals grows very quickly since they are coming from multiple cells. Moreover,  in contrast to natural Wannier orbitals, the natural orbitals present in the $\Gamma$-point calculations are breaking the periodic symmetry of the periodic problem. 
The natural Wannier orbitals $W_j^{0r}$ from Eq.~\ref{eq:natural_wannier_ft} can be complex since they are are constructed with the help of Fourier transform from the k-space to the real space using complex matrices $[U(k)]_j$. Since $W_j^{0r}$ are used to perform the transformation of the one-body and two-body integrals necessary for the impurity problem, the presence of complex $W_j^{0r}$ is problematic since it results in complex transformed integrals.
We observe that in a non-degenerate case, complex $W_j^{0r}$ can be avoided when eigenvectors $ [U(k)]_j$ are not allowed to have arbitrary phase factors equal to $\pm 1$. We eliminate the arbitrary phase factors by multiplying all the elements of each eigenvector by the sign of its first element. 

Finally, let us mention that the density matrix expressed in natural Wannier orbital basis has interesting physical properties~\cite{Koch_Goedecker,Kudinov}. At zero temperature, when expressed in the natural Wannier basis, 
the density matrix $D_{ij}^{rr'}$ decays exponentially with increasing distance between cells $r$ and $r'$. In metallic systems, $D_{ij}^{rr'}$ decays only algebraically due to a discontinuity at the Fermi surface. At finite temperature, the decay of the density matrix becomes exponential for both insulators and metals~\cite{Goedecker}. 

\section{Impurity self-energy and sceening}\label{sec:screening}

The impurity problem in SEET is characterized by two features. First, as already mentioned in Sec.~\ref{sec:algo}, in a stark contrast to the LDA+DMFT procedure, in SEET, which is a fully ab-initio procedure, we do not use any empirical parameters such as empirically chosen effective interactions $U$ and $J$. Second, in SEET, in contrast to the GW+EDMFT method, the impurity interactions are bare Coulomb interactions and are frequency independent. However, it should be stressed that these interactions are not Coulomb interactions in an atomic basis. These two-body interactions, defined as $v^{OR}_{ijkl \in sub}$, are expressed in the orbital basis that was chosen for the impurity problem. Consequently, in our calculations the two-body integrals are transformed to the orthogonal Wannier or natural Wannier basis. 

In the embedding problems, the renormalization of the impurity interactions arises due to carrying out the impurity calculation in a small, truncated orbital subset, while the entire problem is defined in a large orbital space. Consequently, a full self-energy of a strongly correlated orbital subset is expressed as
\begin{equation}\label{eq:embedding_screening}
[\Sigma_{sub}(i\omega)]_{ij}=[\Sigma_{imp}(i\omega)]_{ij}+[\Sigma_{embedding}(i\omega)]_{ij},
\end{equation}
where $[\Sigma_{imp}]_{ij}$ is obtained from the solution of an impurity problem with a set of two-body integrals $v^{OR}_{ijkl \in sub}$.
Independent of the level of theory used to evaluate the term $[\Sigma_{embedding}(i\omega)]_{ij}$, it is calculated using $v^{OR}_{ijkl}$, where at least one of the labels $i,j,k$, or $l$ is not contained in the strongly correlated orbital subset, here denoted as $sub$. This means that effectively, the magnitude of the impurity self-energy $[\Sigma_{imp}(i\omega)]_{ij}$ is adjusted, when the non-local interactions are present,  by the term $[\Sigma_{embedding}(i\omega)]_{ij}$ arising due to the non-local interactions. 

In methods such as LDA+DMFT, the effective interactions $U$ and $J$ are necessary since at the LDA level $[\Sigma_{embedding}(i\omega)]_{ij}=0$. Consequently, in order to recover the proper magnitude of $[\Sigma_{sub}(i\omega)]_{ij}$ which comes from the whole system, not just the impurity problem, the impurity problem has to be reparametrized in such a way that $[\Sigma_{imp}(i\omega)]_{ij}$ evaluated with a new adjusted effective interactions $U$ and $J$, here denoted as $[\Sigma^{eff \ int}_{imp}(i\omega)]_{ij}$, will fulfill 
\begin{equation}
[\Sigma_{sub}(i\omega)]_{ij}=[\Sigma^{eff \ int}_{imp}(i\omega)]_{ij}.
\end{equation}
For a detailed discussion of the interaction renormalization in HF+DMFT or LDA+DMFT see Ref.~\onlinecite{Rusakov14}.

Similarly, in the GW+EDMFT calculations for realistic materials, the parent value of the frequency dependent $W(i\omega)$ is modified in the impurity problem since GW+EDMFT calculations start form a constrained RPA (cRPA) step in which the full physical problem involving all orbitals is mapped onto a smaller problem in which GW is solved. Due to this mapping and then a subsequent adjustment of the impurity problem to retain only the density-density interactions, the  impurity value of $W(i\omega)$ is adjusted in comparison to the parent value of $W_{imp}(i\omega)$ when all the orbitals are present.

In SEET, none of the above mentioned complications arise since all the orbitals are present in the calculation. This means that the term $[\Sigma_{embedding}(i\omega)]_{ij}$ from Eq.~\ref{eq:embedding_screening} is always present and is recovered at the level of the weakly correlated method used in the SEET(method$_\text{stong}$/method$_\text{weak}$) scheme.  
The presence of this term ensures that the high frequency limits of $[\Sigma_{sub}(i\omega)]_{ij}$ and $[\Sigma_{imp}(i\omega)]_{ij}+[\Sigma_{embedding}(i\omega)]_{ij}$ are  equal. 
Consequently, in SEET, the self-energy of a strongly correlated orbital  always contains two terms, a term due to the interactions present among the impurity orbitals which is then ``renormalized'' or ``adjusted'' due to the presence of the second term that arises due to non-local interactions. 
While there is no renormalization of the Coulomb interactions in SEET, the magnitude of the self-energy of the strongly correlated orbitals is renormalized as we explained in detail.

\section{Exact limits and computational scaling}\label{sec:limits_scaling}
\subsection{Computational scaling}

The computational scaling of the SEET algorithm for periodic systems is proportional to the scaling of the weakly correlated method for the whole system plus the scaling of the impurity solver. The impurity solver is run only in the subspace of strongly correlated orbitals or in the subspace of orbitals that require an accurate correlational treatment. This subset of orbitals is present only in the unit cell. Thus, in case of SEET(FCI/GF2), where the GF2 method is used as a weak correlation method, the total computational scaling is $O(n_{\tau}n_{cell}^4 n^5)+O(N_{imp} {{n_s}\choose{n_e}})$. The scaling of a periodic GF2 code is $O(n_{\tau}n_{cell}^4 n^5)$, where $n_{\tau}$ is the number of imaginary time grid points. 
Here, $N_{imp}$ stands for the total number of impurities that can be built among the $n_{act}$ orbitals. The total number of orbitals present in the impurity model is denoted as $n_s=n_{act_{sub}}+n_{b}$, where $n_{act_{sub}}$ is the number of strongly correlated impurity orbitals and $n_b$ is the number of bath orbitals. Here, $n_{act_{sub}}$ stands for the number of strongly correlated orbitals used to build the impurity model. Note that $n_{act_{sub}}$  is usually lesser or equal to $n_{act}$ since strongly correlated orbitals can be divided into multiple impurity problems, for details see Ref.~\onlinecite{Tran_generalized_seet}.

Note that while the scaling of the impurity solver in the case of the full configuration interaction (FCI) or restricted active space configuration interaction (RASCI) is exponential, this expensive step has to be performed only once per iteration in the high level loop. Additionally, $N_{imp}$,  ${n_s}$, and $n_e$ are usually small in comparison to the number of real space cells  $n_{cell}$ or the number of orbitals in each cell $n$.

Consequently, for a realistic periodic problem the total computational scaling of SEET is usually completely dominated by the cost of the weak correlation method. The overall time of the weakly correlated calculation for the whole solid is much longer than the time of the calculation involving the solver capable of treating strong correlations since the solver calculation is  done only for tiny impurity problems. 

\subsection{Exact limits}
In SEET, the impurity orbitals that are treated by an expensive and accurate method are chosen only within the unit cell. This is done to stop the growth of the number of strongly correlated orbitals and make the overall calculations of strongly correlated solids affordable. Consequently, performing SEET(FCI/method$_\text{weak}$) even with all the orbitals in each unit cell treated with the FCI solver, will not result in the FCI cell energy since all the orbitals that do not belong to the unit cell are treated only by an approximate method, here denoted as method$_\text{weak}$=HF, GF2, or GW. 

In cases, where the interaction between multiple unit cells is essential, provided that the computational cost can be handled, the FCI cell energy can be obtained directly or extrapolated to by treating larger and larger supercells. In these supercells,  the increasing number of orbitals would need to be treated by the FCI solver. 
Consequently, SEET has the following exact limits. The limit of vanishing two-body interactions $v_{ijkl}=0$, where SEET(FCI/HF) is equivalent to the non-interacting limit. 
The limit of increasing supercell size, where the true cell energy for a given method within a given basis, can be reach by performing larger and larger supercell calculations using a given method.
Increasing the accuracy of the treatment of the weakly correlated orbitals will also lead to an exact answer within a basis provided that FCI was used to treat the weakly correlated orbitals. 
Finally, if SEET is performed as SEET(method1/method1) where method1 is used both for treating the strongly correlated orbitals in the impurity and simultaneously the same method1 is used to treat the weakly correlated orbitals contained in the environment, then the result of such a treatment is equivalent to a calculation where method1 is used to treat the whole periodic system. For example, SEET(GF2/GF2) is equivalent to performing a GF2 calculation on the whole periodic system.

In practice, for any realistic solids with a large number of orbitals per cell reaching the exact limits which will result in the FCI cell energy is impossible.
Nevertheless, a SEET(FCI/method$_\text{weak}$) calculation, where in a given unit cell strongly correlated orbitals or the orbitals that require an accurate treatment are included in an impurity model calculated by the FCI solver, should result in a very good both qualitative and quantitative approximation to the FCI cell energy. Moreover, all the properties within the unit cell should be in a very good quantitative agreement with the FCI data within a given basis set. 

\section{Results}\label{sec:results}

In this section, we test our implementation of SEET for solids versus established auxiliary filed quantum Monte Carlo (AFQMC) data. These tests allow us to gain understanding of the performance of periodic SEET and compare this finite temperature method to established zero temperature methods.
\subsection{Calibration on 1D hydrogen solid}

\begin{figure} [htb]
  \includegraphics[width=\columnwidth,height=2cm]{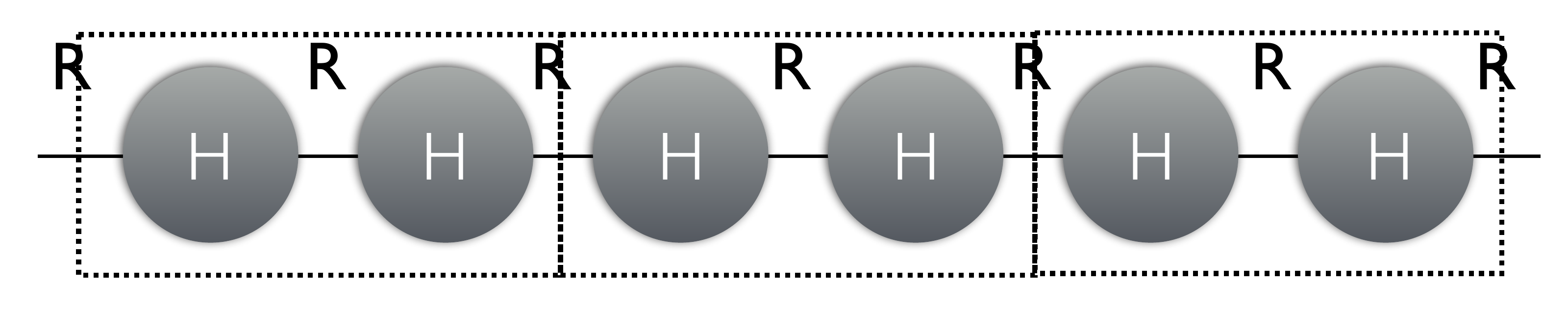} \caption{Two hydrogen atoms are used to define a unit cell in the 1D hydrogen solid. The distance between atoms both within the cell and between cells is same and is denoted as R.}
  \label{fig:hydrogen_cell}
\end{figure}

In the past, in quantum chemistry calculations, hydrogen chains were used frequently as molecular systems in which strong correlation can be modeled~\cite{Hachmann:jcp2006-h50dmrg}. In the same spirit, a 1D hydrogen solid can be used as an example of a periodic system with full Coulomb interactions where the degree of strong correlation can be controlled by changing the distance between hydrogen atoms. 
Here, we assume that in the 1D hydrogen solid, the unit cell is represented by the fragment -(H-H)- due to possible antiferromagnetic ordering. 
Changing the distance $R$ between hydrogens causes adjustment of the distance not only between the unit cells but also within the unit cell, see Fig.~\ref{fig:hydrogen_cell}.

The set of data that we are using to perform the calibration of SEET for periodic problems was obtained by AFQMC and presented in Ref.~\onlinecite{simons_benchmark2}. These AFQMC results were obtained through a series of extrapolations to the thermodynamic limit (TDL) of  the 0 K data for finite hydrogen chains.
In this work, we are using the data obtained in the STO-6G basis set. Since in Ref.~\onlinecite{simons_benchmark2} multiple results were in mutual agreement but were obtained from different methods such as DMRG and AFQMC, we believe that these data reliably represent the ground state of a 1D hydrogen solid within the minimal STO-6G basis. 

To facilitate a comparison with the zero-temperature AFQMC extrapolated to the thermodynamic limit here denoted as AFQMC(0K, TDL), we also evaluated results for periodic RHF, and periodic GF2. 
The RHF result was evaluated at zero-temperature and was converged with respect to k-points and real space cells.
These RHF calculations required 229 k-points and 73 real space cells for  R=1.4 a.u. which was the shortest distance between hydrogen atoms. For R=3.6 a.u., the longest distance between hydrogen atoms, 89 k-points and 29 real space cells were necessary to converge the calculations with respects to both k-space points and real space cells. The RHF calculations have been performed using the \textsc{gaussian}~\cite{g09} program.

The subsequent GF2 calculations use the same number of k-points and real-space cells that is used in RHF. To find a compact GF2 self-energy representation on the imaginary time axis we employed 600 Legendre polynomials~\cite{Kananenka15}. In this paper, we are avoiding building a spline for $G(i\omega)$~\cite{Kananenka16} and we use simple equidistant Matsubara grid containing 30,000 frequencies. In this way both GF2 and SEET Green's functions are built using the same number of frequency points.

Since GF2 is a finite temperature method, the temperature employed had to be low enough so that the ground state of the 1D periodic hydrogen could be recovered and an agreement with the AFQMC(0K, TDL) data can be assured. 
We discovered that when $\beta=20,000$ 1/a.u. corresponding to 15.79 K is used for distances between R=1.4-2.0 a.u.
and $\beta=10,000$ 1/a.u. corresponding to 31.58 K is employed for distances between R=2.4-3.6 a.u., the electronic energy per cell was not changing anymore and our results indicated insulating solutions converged with respect to the value of inverse temperature $\beta$.
Similarly, AFQMC results also indicate that for the distances greater or equal than R=1.4 a.u. only insulating solutions~\cite{} were found at 0 K in the STO-6G basis. While both the SEET and AFQMC results agree here, they also may be an artifact of very slow convergence with the size of the total system. 

In our calculations, we strive to put in the impurity model employed in SEET the most correlated orbitals since such a treatment leads to the best recovery of correlation energy and correlation effects. 
One way of diagnosing how to construct such most correlated orbitals is to observe the magnitude of the hybridization as expressed in Eq.~\ref{hybrid} that is present between the impurity orbitals and the rest of the problem.
We observe that for distances $R\ge 2.0$ a.u., the regular Wannier orbitals described in Sec.~\ref{sec:orbitals} are minimizing the magnitude of the hybridization and lead to better energies. 
For distances $R < 2.0$ a.u. the natural Wannier orbitals, which are delocalized,  lead to superior energies. This can be easily rationalized by noticing that a suitable orbital basis for the somewhat compressed geometries should be built from delocalized orbitals, consequently natural Wannier orbitals are a good basis.

Our reported results use these two different orbital bases for different distances. While in traditional quantum chemistry switching an orbital basis during modeling of a potential energy curve would be frowned upon, here we believe that it is justified since the magnitude of the hybridization provides a diagnostic tool allowing us to predict when to use respective orbital bases. Moreover, here we are only choosing the best definition of orbitals that should be contained in the impurity problem and it is easy to accept that such a definition will be both distance and system dependent.

\begin{figure} [htb]
  \includegraphics[width=\columnwidth]{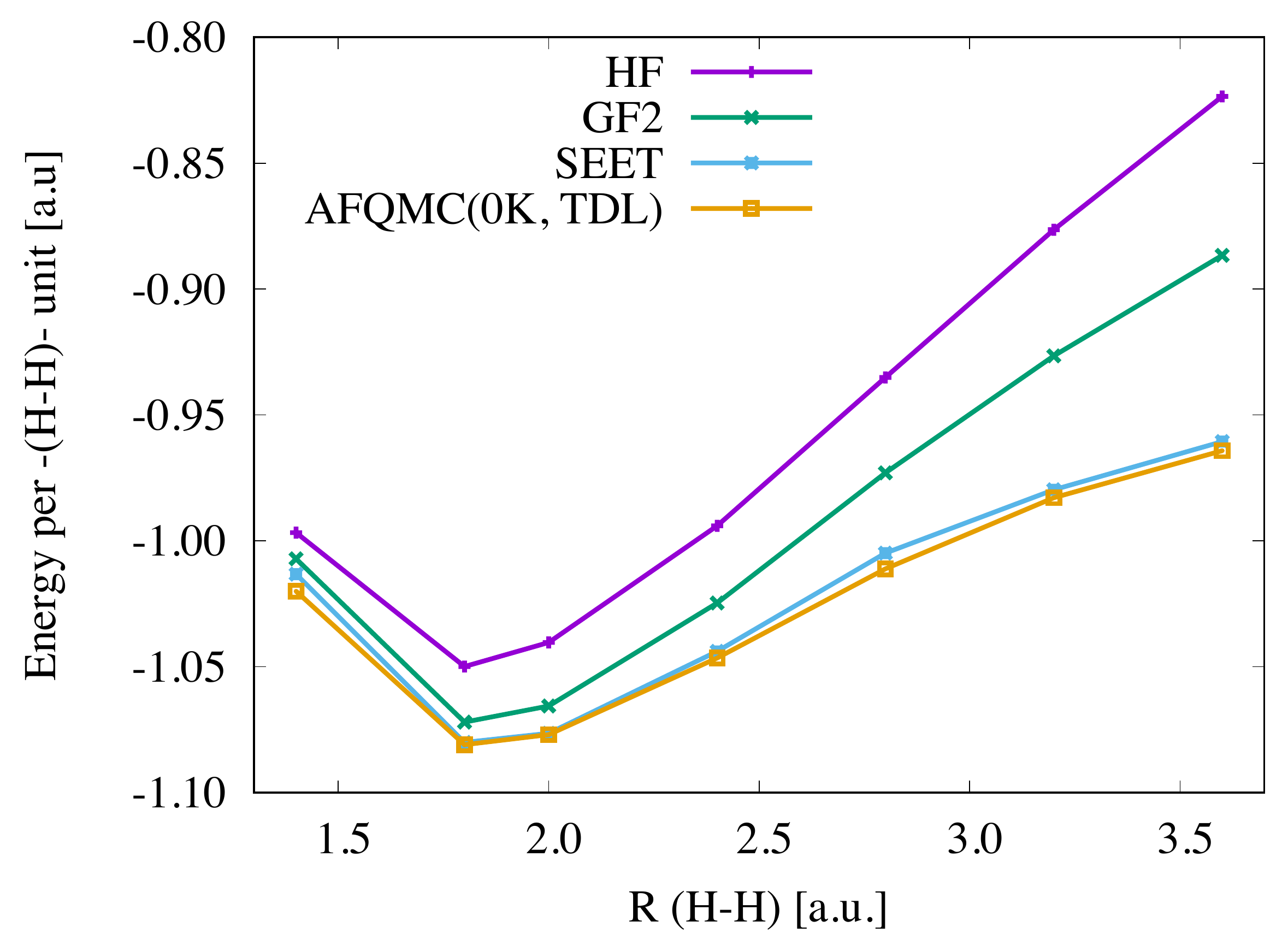} \caption{The total energy per -(H-H)- cell from HF, GF2, SEET, and AFQMC(0K, TDL). Periodic HF is carried out at 0 K. Both GF2 and SEET are evaluated at finite temperatures, for details see the description in the text.}
  \label{fig:seet_gf2_hf_afqmc}
\end{figure}

In Fig.~\ref{fig:seet_gf2_hf_afqmc}, we present a comparison of energy per -(H-H)- cell as a function of inter-hydrogen distance.
As expected, for the points where the inter-hydrogen distance is large, the difference between AFQMC(0K, TDL), HF, and GF2 is large. This difference arises in GF2 built on RHF reference due to to its quantitative failure to describe the Mott phase which requires a description of open-shell singlets.
For RHF this failure is qualitative.

For inter-hydrogen distances below R=2.0 a.u., the agreement between periodic GF2 and AFQMC(0K, TDL) is improved and the differences are around 10 mHa.
For all the distances mentioned, SEET(FCI/GF2)-[2o] is in a very good agreement with AFQMC(0K, TDL) data with the largest difference being 6 mHa for R=1.4 and R=2.8 a.u. All the other differences between SEET(FCI/GF2)-[2o]  and  AFQMC(0K, TDL) are below 3 mHa.
SEET(FCI/GF2)-[2o] stands for a GF2 calculation performed on the whole solid and then a subsequent DMFT-like self-consistency employing a two-orbital impurity Hamiltonian containing -(H-H)- fragment embedded in a bath. The two-orbital impurity Hamiltonian is treated by an FCI solver. 


The most likely reason for the differences between SEET(FCI/GF2)-[2o] and AFQMC(0K, TDL) is the inexactness of SEET(FCI/GF2)-[2o] when only a two-orbital impurity Hamiltonian is treated at the FCI level. This inexactness is the necessary price that is paid for making the SEET(FCI/GF2)-[2o] calculation affordable and limiting the FCI calculation to only impurities containing two orbitals per cell. 

Another possible source of minor differences (less than 1 mHa) lies in possible small errors present in the extrapolation of the AFQMC data to TDL. Additionally, small inaccuracies (smaller than 1 mHa) may appear due the size of the frequency grid which has 30,000 points.

\begin{figure} [htb]
  \includegraphics[width=\columnwidth]{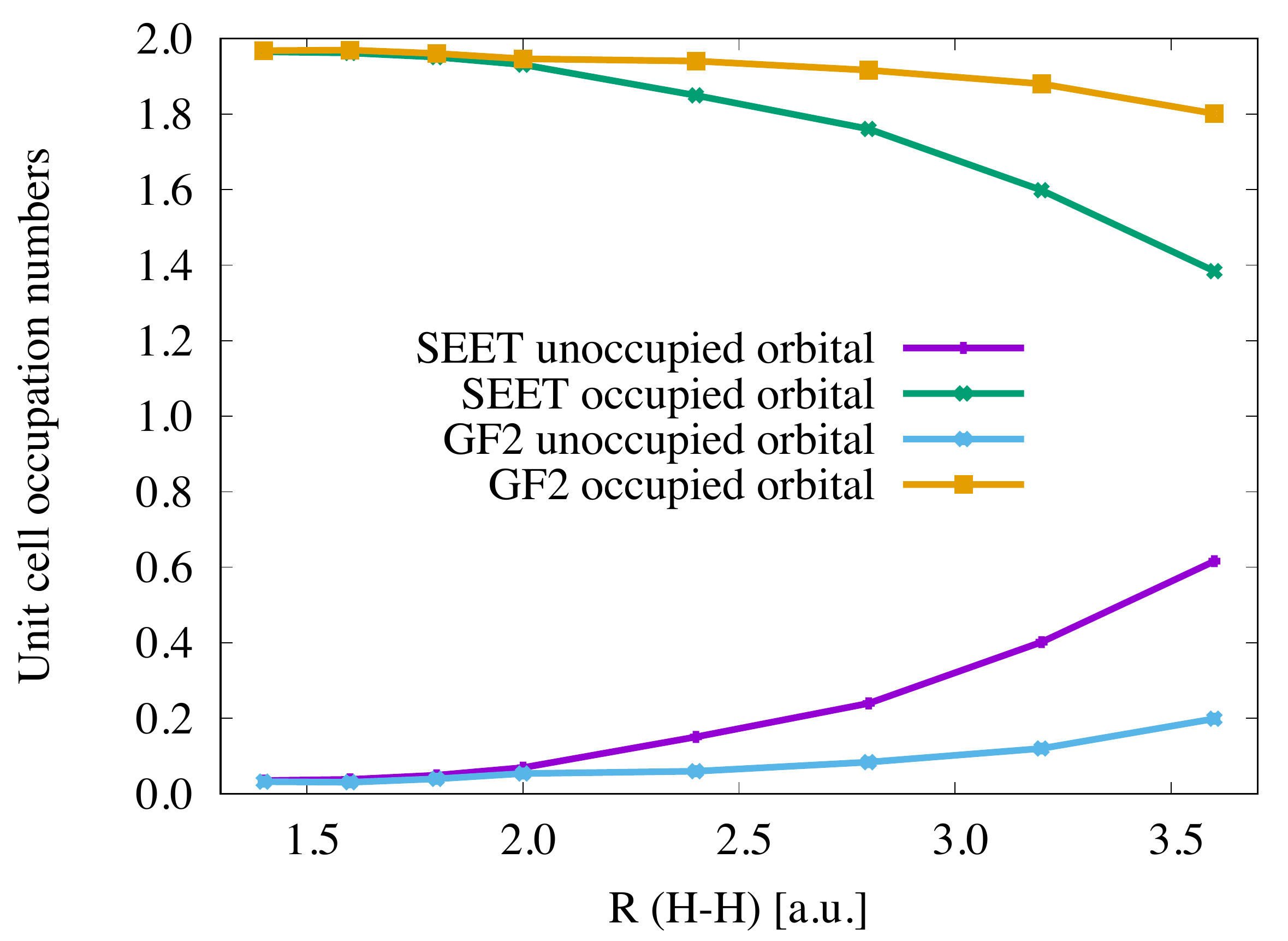} \caption{Occupation numbers in the -(H-H)- unit cell for the 1D hydrogen solid as a function of R(H-H) the interhydrogen distance in a.u.}
  \label{fig:occupation_num_H}
\end{figure}

In Fig.~\ref{fig:occupation_num_H}, we list occupation numbers both from GF2 and SEET(FCI/GF2)-[2o] to illustrate how they change due to performing SEET. For lager inter-hydrogen distances, GF2 natural occupation numbers remain close to 0 and 2 and indicate only a weakly correlated solid. In contrast, the occupation numbers from SEET are far from 0 and 2, remaining closer to 1 and illustrating that for stretched distances the 1D periodic hydrogen is displaying a strongly correlated characteristics, as should be expected. 

\begin{figure*} [htbt]
  \includegraphics[width=\textwidth]{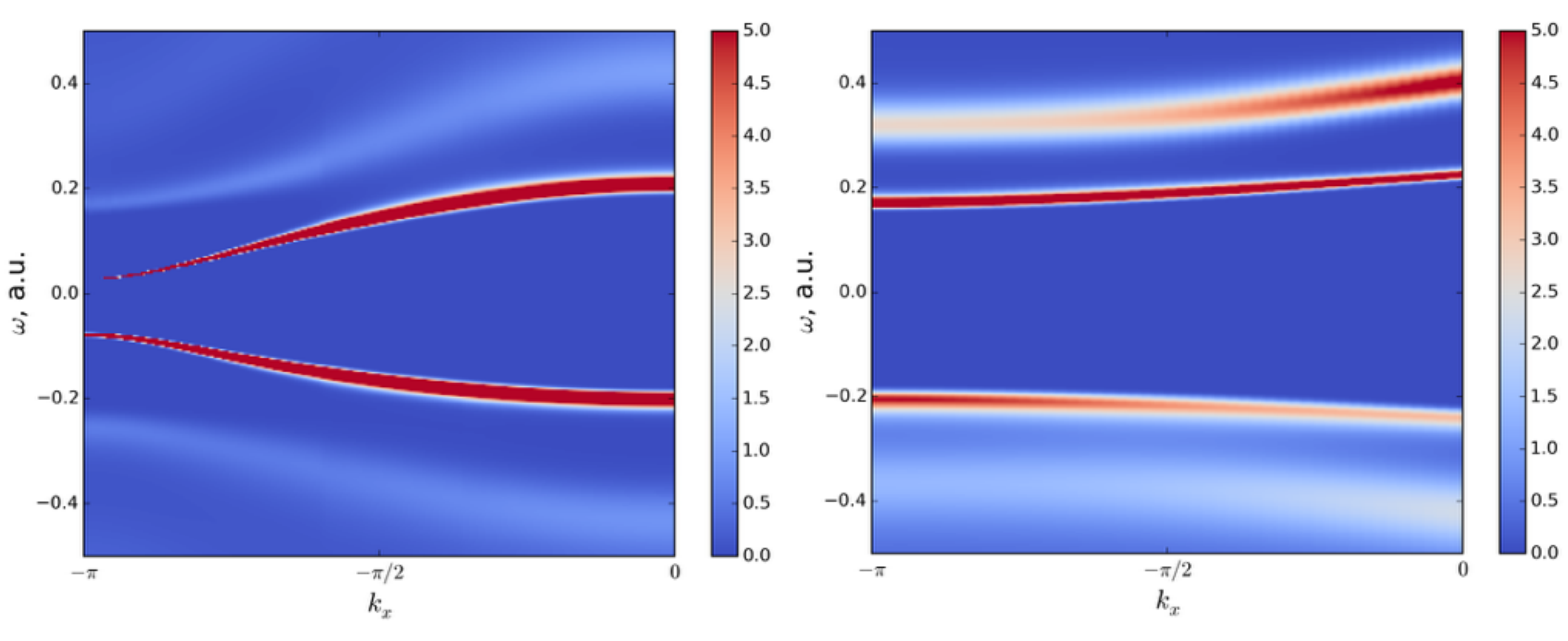} \caption{Left panel: GF2 spectrum for 1D periodic hydrogen for R=3.3 a.u. and $\beta=10000$ 1/a.u.
Right panel: SEET(FCI/GF2)-[2o] spectrum for 1D periodic hydrogen for R=3.3 a.u. and $\beta=10000$  1/a.u.}
  \label{fig:spec_H}
\end{figure*}

Finally, for R=3.3 a.u. in the 1D hydrogen solid, we consider GF2 and SEET photoelectron spectra simulated by analytically continuing the Matsubara Green's function to the real axis via the maximum entropy method.~\cite{LEVY2017149} 
We choose this geometry since for these stretched geometries the difference between the GF2 and SEET spectrum is the most clearly visible as illustrated in Fig.~\ref{fig:spec_H}. For smaller inter-hydrogen distances the GF2 and SEET spectra are displaying only minor differences. 
For R=3.3 a.u., the GF2 spectrum shows an insulating character while both GF2 bands show some variation with respect to k-points typical for a band insulator.  In contrast SEET results show flat bands with almost no variation with respect to k-points. Such a lack of variation is a symptom of an onset of strongly correlated behavior leading to a Mott insulator phase.

\section{Comparison with DMFT}~\label{sec:dmft_comparison}

Here, for periodic problems, we focus on comparison of the SEET scheme  with GW+EDMFT as presented and implemented in Refs.~\onlinecite{PhysRevMaterials.1.043803, PhysRevB.95.245130,PhysRevB.94.201106}. While some of the listed differences are technical and arise due to a particular implementation decision, we still list them since we believe that it is for the benefit of the reader to understand both theoretical and practical side of how both of these methods are executed.

\subsection{Projection to a smaller subspace}

The GW+EDMFT calculation~\cite{PhysRevMaterials.1.043803, PhysRevB.95.245130,PhysRevB.94.201106} starts from the cRPA calculation done on the whole periodic problem. Subsequently a small energy/orbital window, here called sub1, is chosen containing a limited number of orbitals. 
The cRPA construction delivers the renormalized value of screened interactions $W^\text{sub1}(\omega)$ for the orbitals present only in sub1. 
This means that in GW+EDMFT, the fully iterated GW method is only explicitly evaluated  in sub1 which is smaller than the orbital space of the full problem. 
This fully self-consistent GW results in the self-energy, $\Sigma^\text{sub1}_\text{weak}(\omega)$. This also means that the interactions in GW+EDMFT are renormalized in the following sequence $W^\text{full}(\omega) \rightarrow W^\text{sub1}_\text{cRPA}(\omega) \rightarrow W^\text{sub1}_\text{sc-GW}(\omega)$. 
In contrast, we start the SEET calculation by treating with a weakly correlated method, such as self-consistent GF2 or GW, all the orbitals present in the periodic problem. 
This means that the self-energy, $\Sigma^\text{full}_\text{weak}(\omega)$ is always evaluated and present for the whole problem for all orbitals present in the system. Consequently, in SEET there is no renormalization of interactions due to a projection from the full space to a smaller subspace such as sub1. If SEET is evaluated on top of GW, then only $W^\text{full}(\omega)$ interactions are present and these if possible should be evaluated self-consistently. Whenever SEET is using GW or GF2 as weakly correlated method in both cases the full self-energy $\Sigma^\text{full}_\text{weak}(\omega)$ for the whole problem is evaluated. 

\subsection{Choice of the orbital space}

Both of the methods can be executed using many possible orbital bases. Here, we highlight some of the more customary choices that are present in each of the methods.
In GW+EDMFT, a suitable orbital space sub1 is found usually based on DFT calculations which means that a set of strongly correlated orbitals are chosen near the Fermi level by finding a suitable energy window. Alternatively, physical intuition may be used to find the strongly correlated orbitals. It is a common practice that the selected orbitals in the GW+EDMFT method are localized and create a set of maximally localized Wannier orbitals.
As we have mentioned in Sec.~\ref{sec:orb_for_imp}, in SEET to get the best results, the basis in which the strongly correlated orbitals are chosen should minimize the hybridization between the strongly and weakly correlated orbitals.
Thus, depending on the physical situation different orbital bases are possible. We find that in many instances natural periodic orbitals minimize hybridization and in SEET they are obtained from an a priori GW or GF2 calculation yielding a correlated one-body density matrix present in each unit cell.   

\subsection{Impurity problem}

The most pronounced difference between SEET(FCI/GF2 or GW) and GW+EDMFT scheme is present in the treatment of the impurity problem. In SEET, the bare interactions $v$ are used to parametrize the impurity problem. This means that a multitude of existing solvers such as ED, its truncated versions, or hybridization expansion solver can be employed to deliver the impurity Green's function. SEET can be executed with both explicit Hamiltonian and action formulation solvers. In SEET, it is possible to deal with non-diagonal hybridizations since they can be handled successfully by exact diagonalization type solvers. However, when natural Wannier orbitals are employed, the non-diagonal hybridization character is either minimized or avoided completely.
In SEET, we never introduce any approximation to the Hamiltonian and consequently, we do not modify the structure of the two-body interactions. Such a modification is not necessary in SEET since solvers such as ED or truncated ED can treat the full two-body interaction $v$ tensor.
In SEET, it is also possible to employ its generalized version and use intersecting impurity problems that lead to the overall SEET functional in the form of Eq.~\ref{eq:seet_mix_func}.
In the GW+EDMFT method, the impurity problem is solved using the retarded interactions, thus at present only the action formulation solvers can be employed to yield the impurity Green's function. Consequently, current solver choices are restricted to the hybridization expansion or continuous time auxiliary field quantum Monte Carlo since these solvers are capable of dealing with frequency dependent interactions. One practical execution problem appearing due to these solvers is a sign problem arising in case of non-diagonal hybridizations. Thus, frequently, in practical calculations to minimize the sign problem off-diagonal elements of the hybridization matrix are neglected. Moreover, again to minimize the possible sign problem, frequently only density-density elements of the two-body frequency dependent integrals are used in practical calculations.

\section{Conclusions}\label{sec:conclusions}

We have presented an extension of the SEET algorithm that is suitable for periodic systems. In this algorithm, as in SEET for molecular systems, we 
choose the impurity problems in a basis that maximally decouples the weakly and strongly correlated orbitals and minimizes the hybridization between the strongly correlated orbitals present in the impurity and the remaining weakly correlated ones. A particular choice of such a basis is problem dependent and here we have focused on the discussion of natural Wannier orbital basis. In such a basis the two-body interactions are non-local and the orbitals are delocalized. The impurity problem is solved using bare interactions. These interactions are obtained by a basis transform of all the interactions present in the system to the basis that minimizes hybridization between the strongly and weakly correlated orbitals. 

We have contrasted these features of SEET with the features of the GW+EDMFT theory. The most stark difference between these approaches is present in the diagrammatic form of the Luttinger-Ward functional resulting from the lack of screened interactions present in the impurity problem. This means in case of SEET(FCI/GF2) that if any diagram containing two-body interaction has at least one label that is not in the impurity then this diagram has only renormalized Green's function lines and not interactions. In case of SEET(FCI/GW), any diagram containing two-body interactions that  has at least one label that is not in the impurity then this diagram has renormalized Green's function lines and interactions, however, the renormalization of the interactions is only at the GW level.

To ensure that periodic SEET(FCI/GF2) can reach high accuracy, we compared it to AFQMC evaluated at 0K and extrapolated to the thermodynamic limit for finite hydrogen chain sizes. SEET was performed on a 1D hydrogen solid and to reach an agreement with AFQMC at 0K, the temperature in SEET had to be lowered to 15-30 K depending on the lattice spacing. We demonstrated that for the regimes considered, SEET(FCI/GF2) is recovering the AFQMC energies to a very good accuracy using only a single impurity that encompasses a single unit cell, here -(H-H)-.

While this work demonstrates the potential of embedding methods such as SEET to reach a high accuracy in periodic systems with an expensive treatment restricted only to the unit cell orbitals, more cases will need to be analyzed in the future to fully understand its advantages and limitations. 
Here, a particular focus should be placed on establishing classes of systems and regimes for which SEET can give good results. Simultaneously, investigations should be carried out to establish in which materials and regimes SEET and GW+EDMFT will display significant differences.

\section{Acknowledgements}

D.Z. and A.A.R. would like to acknowledge support from the US Department of Energy (DOE) grant No. ER16391. S.I. and L.N.T. are acknowledging support by the Simons Foundation via the Simons Collaboration on the Many-Electron Problem.
D.Z. thanks the Simons Foundation for the sabbatical support. The Flatiron Institute is a division of the Simons Foundation.
D.Z. also would like to thank Emanuel Gull for reading this manuscript.


\begin{thebibliography}{52}%
\makeatletter
\providecommand \@ifxundefined [1]{%
 \@ifx{#1\undefined}
}%
\providecommand \@ifnum [1]{%
 \ifnum #1\expandafter \@firstoftwo
 \else \expandafter \@secondoftwo
 \fi
}%
\providecommand \@ifx [1]{%
 \ifx #1\expandafter \@firstoftwo
 \else \expandafter \@secondoftwo
 \fi
}%
\providecommand \natexlab [1]{#1}%
\providecommand \enquote  [1]{``#1''}%
\providecommand \bibnamefont  [1]{#1}%
\providecommand \bibfnamefont [1]{#1}%
\providecommand \citenamefont [1]{#1}%
\providecommand \href@noop [0]{\@secondoftwo}%
\providecommand \href [0]{\begingroup \@sanitize@url \@href}%
\providecommand \@href[1]{\@@startlink{#1}\@@href}%
\providecommand \@@href[1]{\endgroup#1\@@endlink}%
\providecommand \@sanitize@url [0]{\catcode `\\12\catcode `\$12\catcode
  `\&12\catcode `\#12\catcode `\^12\catcode `\_12\catcode `\%12\relax}%
\providecommand \@@startlink[1]{}%
\providecommand \@@endlink[0]{}%
\providecommand \url  [0]{\begingroup\@sanitize@url \@url }%
\providecommand \@url [1]{\endgroup\@href {#1}{\urlprefix }}%
\providecommand \urlprefix  [0]{URL }%
\providecommand \Eprint [0]{\href }%
\providecommand \doibase [0]{http://dx.doi.org/}%
\providecommand \selectlanguage [0]{\@gobble}%
\providecommand \bibinfo  [0]{\@secondoftwo}%
\providecommand \bibfield  [0]{\@secondoftwo}%
\providecommand \translation [1]{[#1]}%
\providecommand \BibitemOpen [0]{}%
\providecommand \bibitemStop [0]{}%
\providecommand \bibitemNoStop [0]{.\EOS\space}%
\providecommand \EOS [0]{\spacefactor3000\relax}%
\providecommand \BibitemShut  [1]{\csname bibitem#1\endcsname}%
\let\auto@bib@innerbib\@empty
\bibitem [{\citenamefont {Kohn}\ and\ \citenamefont {Sham}(1965)}]{Kohn65}%
  \BibitemOpen
  \bibfield  {author} {\bibinfo {author} {\bibfnamefont {W.}~\bibnamefont
  {Kohn}}\ and\ \bibinfo {author} {\bibfnamefont {L.~J.}\ \bibnamefont
  {Sham}},\ }\href {\doibase 10.1103/PhysRev.140.A1133} {\bibfield  {journal}
  {\bibinfo  {journal} {Phys. Rev.}\ }\textbf {\bibinfo {volume} {140}},\
  \bibinfo {pages} {A1133} (\bibinfo {year} {1965})}\BibitemShut {NoStop}%
\bibitem [{\citenamefont {Georges}\ \emph {et~al.}(1996)\citenamefont
  {Georges}, \citenamefont {Kotliar}, \citenamefont {Krauth},\ and\
  \citenamefont {Rozenberg}}]{Georges96}%
  \BibitemOpen
  \bibfield  {author} {\bibinfo {author} {\bibfnamefont {A.}~\bibnamefont
  {Georges}}, \bibinfo {author} {\bibfnamefont {G.}~\bibnamefont {Kotliar}},
  \bibinfo {author} {\bibfnamefont {W.}~\bibnamefont {Krauth}}, \ and\ \bibinfo
  {author} {\bibfnamefont {M.~J.}\ \bibnamefont {Rozenberg}},\ }\href {\doibase
  10.1103/RevModPhys.68.13} {\bibfield  {journal} {\bibinfo  {journal} {Rev.
  Mod. Phys.}\ }\textbf {\bibinfo {volume} {68}},\ \bibinfo {pages} {13}
  (\bibinfo {year} {1996})}\BibitemShut {NoStop}%
\bibitem [{\citenamefont {Georges}\ and\ \citenamefont
  {Kotliar}(1992)}]{Georges92}%
  \BibitemOpen
  \bibfield  {author} {\bibinfo {author} {\bibfnamefont {A.}~\bibnamefont
  {Georges}}\ and\ \bibinfo {author} {\bibfnamefont {G.}~\bibnamefont
  {Kotliar}},\ }\href {\doibase 10.1103/PhysRevB.45.6479} {\bibfield  {journal}
  {\bibinfo  {journal} {Phys. Rev. B}\ }\textbf {\bibinfo {volume} {45}},\
  \bibinfo {pages} {6479} (\bibinfo {year} {1992})}\BibitemShut {NoStop}%
\bibitem [{\citenamefont {Georges}(2004)}]{Georges04}%
  \BibitemOpen
  \bibfield  {author} {\bibinfo {author} {\bibfnamefont {A.}~\bibnamefont
  {Georges}},\ }\href {\doibase http://dx.doi.org/10.1063/1.1800733} {\bibfield
   {journal} {\bibinfo  {journal} {AIP Conference Proceedings}\ }\textbf
  {\bibinfo {volume} {715}},\ \bibinfo {pages} {3} (\bibinfo {year}
  {2004})}\BibitemShut {NoStop}%
\bibitem [{\citenamefont {Biermann}\ \emph {et~al.}(2003)\citenamefont
  {Biermann}, \citenamefont {Aryasetiawan},\ and\ \citenamefont
  {Georges}}]{Biermann03}%
  \BibitemOpen
  \bibfield  {author} {\bibinfo {author} {\bibfnamefont {S.}~\bibnamefont
  {Biermann}}, \bibinfo {author} {\bibfnamefont {F.}~\bibnamefont
  {Aryasetiawan}}, \ and\ \bibinfo {author} {\bibfnamefont {A.}~\bibnamefont
  {Georges}},\ }\href {\doibase 10.1103/PhysRevLett.90.086402} {\bibfield
  {journal} {\bibinfo  {journal} {Phys. Rev. Lett.}\ }\textbf {\bibinfo
  {volume} {90}},\ \bibinfo {pages} {086402} (\bibinfo {year}
  {2003})}\BibitemShut {NoStop}%
\bibitem [{\citenamefont {Biermann}\ \emph {et~al.}(2005)\citenamefont
  {Biermann}, \citenamefont {Aryasetiawan},\ and\ \citenamefont
  {Georges}}]{Biermann05}%
  \BibitemOpen
  \bibfield  {author} {\bibinfo {author} {\bibfnamefont {S.}~\bibnamefont
  {Biermann}}, \bibinfo {author} {\bibfnamefont {F.}~\bibnamefont
  {Aryasetiawan}}, \ and\ \bibinfo {author} {\bibfnamefont {A.}~\bibnamefont
  {Georges}},\ }\enquote {\bibinfo {title} {Electronic structure of strongly
  correlated materials: Towards a first principles scheme},}\ in\ \href
  {\doibase 10.1007/1-4020-2708-7_4} {\emph {\bibinfo {booktitle} {Physics of
  Spin in Solids: Materials, Methods and Applications}}},\ \bibinfo {editor}
  {edited by\ \bibinfo {editor} {\bibfnamefont {S.}~\bibnamefont {Halilov}}}\
  (\bibinfo  {publisher} {Springer Netherlands},\ \bibinfo {address}
  {Dordrecht},\ \bibinfo {year} {2005})\ pp.\ \bibinfo {pages}
  {43--65}\BibitemShut {NoStop}%
\bibitem [{\citenamefont {Werner}\ and\ \citenamefont
  {Casula}(2016)}]{GW_review_werner2016}%
  \BibitemOpen
  \bibfield  {author} {\bibinfo {author} {\bibfnamefont {P.}~\bibnamefont
  {Werner}}\ and\ \bibinfo {author} {\bibfnamefont {M.}~\bibnamefont
  {Casula}},\ }\href {http://stacks.iop.org/0953-8984/28/i=38/a=383001}
  {\bibfield  {journal} {\bibinfo  {journal} {Journal of Physics: Condensed
  Matter}\ }\textbf {\bibinfo {volume} {28}},\ \bibinfo {pages} {383001}
  (\bibinfo {year} {2016})}\BibitemShut {NoStop}%
\bibitem [{\citenamefont {Knizia}\ and\ \citenamefont
  {Chan}(2012)}]{dmet_knizia12}%
  \BibitemOpen
  \bibfield  {author} {\bibinfo {author} {\bibfnamefont {G.}~\bibnamefont
  {Knizia}}\ and\ \bibinfo {author} {\bibfnamefont {G.~K.-L.}\ \bibnamefont
  {Chan}},\ }\href {\doibase 10.1103/PhysRevLett.109.186404} {\bibfield
  {journal} {\bibinfo  {journal} {Phys. Rev. Lett.}\ }\textbf {\bibinfo
  {volume} {109}},\ \bibinfo {pages} {186404} (\bibinfo {year}
  {2012})}\BibitemShut {NoStop}%
\bibitem [{\citenamefont {Knizia}\ and\ \citenamefont
  {Chan}(2013)}]{dmet_knizia13}%
  \BibitemOpen
  \bibfield  {author} {\bibinfo {author} {\bibfnamefont {G.}~\bibnamefont
  {Knizia}}\ and\ \bibinfo {author} {\bibfnamefont {G.~K.-L.}\ \bibnamefont
  {Chan}},\ }\href {\doibase 10.1021/ct301044e} {\bibfield  {journal} {\bibinfo
   {journal} {Journal of Chemical Theory and Computation}\ }\textbf {\bibinfo
  {volume} {9}},\ \bibinfo {pages} {1428} (\bibinfo {year} {2013})},\ \bibinfo
  {note} {pMID: 26587604},\ \Eprint
  {http://arxiv.org/abs/http://dx.doi.org/10.1021/ct301044e}
  {http://dx.doi.org/10.1021/ct301044e} \BibitemShut {NoStop}%
\bibitem [{\citenamefont {Welborn}\ \emph {et~al.}(2016)\citenamefont
  {Welborn}, \citenamefont {Tsuchimochi},\ and\ \citenamefont
  {Voorhis}}]{DMET_bootstrap_jcp_2016}%
  \BibitemOpen
  \bibfield  {author} {\bibinfo {author} {\bibfnamefont {M.}~\bibnamefont
  {Welborn}}, \bibinfo {author} {\bibfnamefont {T.}~\bibnamefont
  {Tsuchimochi}}, \ and\ \bibinfo {author} {\bibfnamefont {T.~V.}\ \bibnamefont
  {Voorhis}},\ }\href {\doibase 10.1063/1.4960986} {\bibfield  {journal}
  {\bibinfo  {journal} {The Journal of Chemical Physics}\ }\textbf {\bibinfo
  {volume} {145}},\ \bibinfo {pages} {074102} (\bibinfo {year} {2016})},\
  \Eprint {http://arxiv.org/abs/http://dx.doi.org/10.1063/1.4960986}
  {http://dx.doi.org/10.1063/1.4960986} \BibitemShut {NoStop}%
\bibitem [{\citenamefont {Kananenka}\ \emph {et~al.}(2015)\citenamefont
  {Kananenka}, \citenamefont {Gull},\ and\ \citenamefont {Zgid}}]{Zgid15}%
  \BibitemOpen
  \bibfield  {author} {\bibinfo {author} {\bibfnamefont {A.~A.}\ \bibnamefont
  {Kananenka}}, \bibinfo {author} {\bibfnamefont {E.}~\bibnamefont {Gull}}, \
  and\ \bibinfo {author} {\bibfnamefont {D.}~\bibnamefont {Zgid}},\ }\href
  {\doibase 10.1103/PhysRevB.91.121111} {\bibfield  {journal} {\bibinfo
  {journal} {Phys. Rev. B}\ }\textbf {\bibinfo {volume} {91}},\ \bibinfo
  {pages} {121111} (\bibinfo {year} {2015})}\BibitemShut {NoStop}%
\bibitem [{\citenamefont {Lan}\ \emph {et~al.}(2015)\citenamefont {Lan},
  \citenamefont {Kananenka},\ and\ \citenamefont {Zgid}}]{Tran15b}%
  \BibitemOpen
  \bibfield  {author} {\bibinfo {author} {\bibfnamefont {T.~N.}\ \bibnamefont
  {Lan}}, \bibinfo {author} {\bibfnamefont {A.~A.}\ \bibnamefont {Kananenka}},
  \ and\ \bibinfo {author} {\bibfnamefont {D.}~\bibnamefont {Zgid}},\ }\href
  {\doibase http://dx.doi.org/10.1063/1.4938562} {\bibfield  {journal}
  {\bibinfo  {journal} {The Journal of Chemical Physics}\ }\textbf {\bibinfo
  {volume} {143}},\ \bibinfo {eid} {241102} (\bibinfo {year} {2015}),\
  http://dx.doi.org/10.1063/1.4938562}\BibitemShut {NoStop}%
\bibitem [{\citenamefont {Nguyen~Lan}\ \emph {et~al.}(2016)\citenamefont
  {Nguyen~Lan}, \citenamefont {Kananenka},\ and\ \citenamefont
  {Zgid}}]{Tran16}%
  \BibitemOpen
  \bibfield  {author} {\bibinfo {author} {\bibfnamefont {T.}~\bibnamefont
  {Nguyen~Lan}}, \bibinfo {author} {\bibfnamefont {A.~A.}\ \bibnamefont
  {Kananenka}}, \ and\ \bibinfo {author} {\bibfnamefont {D.}~\bibnamefont
  {Zgid}},\ }\href {http://dx.doi.org/10.1021/acs.jctc.6b00638} {\bibfield
  {journal} {\bibinfo  {journal} {Journal of Chemical Theory and Computation}\
  }\textbf {\bibinfo {volume} {12}},\ \bibinfo {pages} {4856} (\bibinfo {year}
  {2016})}\BibitemShut {NoStop}%
\bibitem [{\citenamefont {Zgid}\ and\ \citenamefont {Gull}(2017)}]{zgid_njp17}%
  \BibitemOpen
  \bibfield  {author} {\bibinfo {author} {\bibfnamefont {D.}~\bibnamefont
  {Zgid}}\ and\ \bibinfo {author} {\bibfnamefont {E.}~\bibnamefont {Gull}},\
  }\href {http://stacks.iop.org/1367-2630/19/i=2/a=023047} {\bibfield
  {journal} {\bibinfo  {journal} {New Journal of Physics}\ }\textbf {\bibinfo
  {volume} {19}},\ \bibinfo {pages} {023047} (\bibinfo {year}
  {2017})}\BibitemShut {NoStop}%
\bibitem [{\citenamefont {Lan}\ and\ \citenamefont
  {Zgid}(2017)}]{Tran_generalized_seet}%
  \BibitemOpen
  \bibfield  {author} {\bibinfo {author} {\bibfnamefont {T.~N.}\ \bibnamefont
  {Lan}}\ and\ \bibinfo {author} {\bibfnamefont {D.}~\bibnamefont {Zgid}},\
  }\href {\doibase 10.1021/acs.jpclett.7b00689} {\bibfield  {journal} {\bibinfo
   {journal} {The Journal of Physical Chemistry Letters}\ }\textbf {\bibinfo
  {volume} {8}},\ \bibinfo {pages} {2200} (\bibinfo {year} {2017})},\ \bibinfo
  {note} {pMID: 28453934},\ \Eprint
  {http://arxiv.org/abs/http://dx.doi.org/10.1021/acs.jpclett.7b00689}
  {http://dx.doi.org/10.1021/acs.jpclett.7b00689} \BibitemShut {NoStop}%
\bibitem [{\citenamefont {Lan}\ \emph {et~al.}(2017)\citenamefont {Lan},
  \citenamefont {Shee}, \citenamefont {Li}, \citenamefont {Gull},\ and\
  \citenamefont {Zgid}}]{Tran_GW_SEET}%
  \BibitemOpen
  \bibfield  {author} {\bibinfo {author} {\bibfnamefont {T.~N.}\ \bibnamefont
  {Lan}}, \bibinfo {author} {\bibfnamefont {A.}~\bibnamefont {Shee}}, \bibinfo
  {author} {\bibfnamefont {J.}~\bibnamefont {Li}}, \bibinfo {author}
  {\bibfnamefont {E.}~\bibnamefont {Gull}}, \ and\ \bibinfo {author}
  {\bibfnamefont {D.}~\bibnamefont {Zgid}},\ }\href {\doibase
  10.1103/PhysRevB.96.155106} {\bibfield  {journal} {\bibinfo  {journal} {Phys.
  Rev. B}\ }\textbf {\bibinfo {volume} {96}},\ \bibinfo {pages} {155106}
  (\bibinfo {year} {2017})}\BibitemShut {NoStop}%
\bibitem [{\citenamefont {Motta}\ \emph {et~al.}(2017)\citenamefont {Motta},
  \citenamefont {Ceperley}, \citenamefont {Chan}, \citenamefont {Gomez},
  \citenamefont {Gull}, \citenamefont {Guo}, \citenamefont {Jim\'enez-Hoyos},
  \citenamefont {Lan}, \citenamefont {Li}, \citenamefont {Ma}, \citenamefont
  {Millis}, \citenamefont {Prokof'ev}, \citenamefont {Ray}, \citenamefont
  {Scuseria}, \citenamefont {Sorella}, \citenamefont {Stoudenmire},
  \citenamefont {Sun}, \citenamefont {Tupitsyn}, \citenamefont {White},
  \citenamefont {Zgid},\ and\ \citenamefont {Zhang}}]{simons_benchmark2}%
  \BibitemOpen
  \bibfield  {author} {\bibinfo {author} {\bibfnamefont {M.}~\bibnamefont
  {Motta}}, \bibinfo {author} {\bibfnamefont {D.~M.}\ \bibnamefont {Ceperley}},
  \bibinfo {author} {\bibfnamefont {G.~K.-L.}\ \bibnamefont {Chan}}, \bibinfo
  {author} {\bibfnamefont {J.~A.}\ \bibnamefont {Gomez}}, \bibinfo {author}
  {\bibfnamefont {E.}~\bibnamefont {Gull}}, \bibinfo {author} {\bibfnamefont
  {S.}~\bibnamefont {Guo}}, \bibinfo {author} {\bibfnamefont {C.~A.}\
  \bibnamefont {Jim\'enez-Hoyos}}, \bibinfo {author} {\bibfnamefont {T.~N.}\
  \bibnamefont {Lan}}, \bibinfo {author} {\bibfnamefont {J.}~\bibnamefont
  {Li}}, \bibinfo {author} {\bibfnamefont {F.}~\bibnamefont {Ma}}, \bibinfo
  {author} {\bibfnamefont {A.~J.}\ \bibnamefont {Millis}}, \bibinfo {author}
  {\bibfnamefont {N.~V.}\ \bibnamefont {Prokof'ev}}, \bibinfo {author}
  {\bibfnamefont {U.}~\bibnamefont {Ray}}, \bibinfo {author} {\bibfnamefont
  {G.~E.}\ \bibnamefont {Scuseria}}, \bibinfo {author} {\bibfnamefont
  {S.}~\bibnamefont {Sorella}}, \bibinfo {author} {\bibfnamefont {E.~M.}\
  \bibnamefont {Stoudenmire}}, \bibinfo {author} {\bibfnamefont
  {Q.}~\bibnamefont {Sun}}, \bibinfo {author} {\bibfnamefont {I.~S.}\
  \bibnamefont {Tupitsyn}}, \bibinfo {author} {\bibfnamefont {S.~R.}\
  \bibnamefont {White}}, \bibinfo {author} {\bibfnamefont {D.}~\bibnamefont
  {Zgid}}, \ and\ \bibinfo {author} {\bibfnamefont {S.}~\bibnamefont {Zhang}}
  (\bibinfo {collaboration} {Simons Collaboration on the Many-Electron
  Problem}),\ }\href {\doibase 10.1103/PhysRevX.7.031059} {\bibfield  {journal}
  {\bibinfo  {journal} {Phys. Rev. X}\ }\textbf {\bibinfo {volume} {7}},\
  \bibinfo {pages} {031059} (\bibinfo {year} {2017})}\BibitemShut {NoStop}%
\bibitem [{\citenamefont {Tran}\ \emph {et~al.}(2018)\citenamefont {Tran},
  \citenamefont {Iskakov},\ and\ \citenamefont {Zgid}}]{Tran_useet}%
  \BibitemOpen
  \bibfield  {author} {\bibinfo {author} {\bibfnamefont {L.~N.}\ \bibnamefont
  {Tran}}, \bibinfo {author} {\bibfnamefont {S.}~\bibnamefont {Iskakov}}, \
  and\ \bibinfo {author} {\bibfnamefont {D.}~\bibnamefont {Zgid}},\ }\href
  {\doibase 10.1021/acs.jpclett.8b01754} {\bibfield  {journal} {\bibinfo
  {journal} {The Journal of Physical Chemistry Letters}\ }\textbf {\bibinfo
  {volume} {9}},\ \bibinfo {pages} {4444} (\bibinfo {year} {2018})},\ \bibinfo
  {note} {pMID: 30024163}\BibitemShut {NoStop}%
\bibitem [{\citenamefont {Roos}(1987)}]{roos1987complete}%
  \BibitemOpen
  \bibfield  {author} {\bibinfo {author} {\bibfnamefont {B.~O.}\ \bibnamefont
  {Roos}},\ }\href@noop {} {\bibfield  {journal} {\bibinfo  {journal} {Advances
  in Chemical Physics: Ab Initio Methods in Quantum Chemistry Part 2, Volume
  69}\ ,\ \bibinfo {pages} {399}} (\bibinfo {year} {1987})}\BibitemShut
  {NoStop}%
\bibitem [{\citenamefont {Andersson}\ \emph {et~al.}(1990)\citenamefont
  {Andersson}, \citenamefont {Malmqvist}, \citenamefont {Roos}, \citenamefont
  {Sadlej},\ and\ \citenamefont {Wolinski}}]{Kerstin:jpc/94/5483}%
  \BibitemOpen
  \bibfield  {author} {\bibinfo {author} {\bibfnamefont {K.}~\bibnamefont
  {Andersson}}, \bibinfo {author} {\bibfnamefont {P.~A.}\ \bibnamefont
  {Malmqvist}}, \bibinfo {author} {\bibfnamefont {B.~O.}\ \bibnamefont {Roos}},
  \bibinfo {author} {\bibfnamefont {A.~J.}\ \bibnamefont {Sadlej}}, \ and\
  \bibinfo {author} {\bibfnamefont {K.}~\bibnamefont {Wolinski}},\ }\href@noop
  {} {\bibfield  {journal} {\bibinfo  {journal} {J. Phys. Chem.}\ }\textbf
  {\bibinfo {volume} {94}},\ \bibinfo {pages} {5483} (\bibinfo {year}
  {1990})}\BibitemShut {NoStop}%
\bibitem [{\citenamefont {Angeli}\ \emph {et~al.}(2002)\citenamefont {Angeli},
  \citenamefont {Cimiraglia},\ and\ \citenamefont {Malrieu}}]{Angeli1}%
  \BibitemOpen
  \bibfield  {author} {\bibinfo {author} {\bibfnamefont {C.}~\bibnamefont
  {Angeli}}, \bibinfo {author} {\bibfnamefont {R.}~\bibnamefont {Cimiraglia}},
  \ and\ \bibinfo {author} {\bibfnamefont {J.-P.}\ \bibnamefont {Malrieu}},\
  }\href
  {http://scitation.aip.org/content/aip/journal/jcp/117/20/10.1063/1.1515317}
  {\bibfield  {journal} {\bibinfo  {journal} {J. Chem. Phys.}\ }\textbf
  {\bibinfo {volume} {117}},\ \bibinfo {pages} {9138} (\bibinfo {year}
  {2002})}\BibitemShut {NoStop}%
\bibitem [{\citenamefont {Angeli}\ \emph {et~al.}(2001)\citenamefont {Angeli},
  \citenamefont {Cimiraglia}, \citenamefont {Evangelisti}, \citenamefont
  {Leininger},\ and\ \citenamefont {Malrieu}}]{Angeli2}%
  \BibitemOpen
  \bibfield  {author} {\bibinfo {author} {\bibfnamefont {C.}~\bibnamefont
  {Angeli}}, \bibinfo {author} {\bibfnamefont {R.}~\bibnamefont {Cimiraglia}},
  \bibinfo {author} {\bibfnamefont {S.}~\bibnamefont {Evangelisti}}, \bibinfo
  {author} {\bibfnamefont {T.}~\bibnamefont {Leininger}}, \ and\ \bibinfo
  {author} {\bibfnamefont {J.-P.}\ \bibnamefont {Malrieu}},\ }\href
  {http://scitation.aip.org/content/aip/journal/jcp/114/23/10.1063/1.1361246}
  {\bibfield  {journal} {\bibinfo  {journal} {J. Chem. Phys.}\ }\textbf
  {\bibinfo {volume} {114}},\ \bibinfo {pages} {10252} (\bibinfo {year}
  {2001})}\BibitemShut {NoStop}%
\bibitem [{\citenamefont {Luttinger}\ and\ \citenamefont
  {Ward}(1960)}]{Luttinger60}%
  \BibitemOpen
  \bibfield  {author} {\bibinfo {author} {\bibfnamefont {J.~M.}\ \bibnamefont
  {Luttinger}}\ and\ \bibinfo {author} {\bibfnamefont {J.~C.}\ \bibnamefont
  {Ward}},\ }\href {\doibase 10.1103/PhysRev.118.1417} {\bibfield  {journal}
  {\bibinfo  {journal} {Phys. Rev.}\ }\textbf {\bibinfo {volume} {118}},\
  \bibinfo {pages} {1417} (\bibinfo {year} {1960})}\BibitemShut {NoStop}%
\bibitem [{\citenamefont {Vu\ifmmode \check{c}\else \v{c}\fi{}i\ifmmode
  \check{c}\else \v{c}\fi{}evi\ifmmode~\acute{c}\else \'{c}\fi{}}\ \emph
  {et~al.}(2018)\citenamefont {Vu\ifmmode \check{c}\else \v{c}\fi{}i\ifmmode
  \check{c}\else \v{c}\fi{}evi\ifmmode~\acute{c}\else \'{c}\fi{}},
  \citenamefont {Wentzell}, \citenamefont {Ferrero},\ and\ \citenamefont
  {Parcollet}}]{PhysRevB.97.125141}%
  \BibitemOpen
  \bibfield  {author} {\bibinfo {author} {\bibfnamefont {J.}~\bibnamefont
  {Vu\ifmmode \check{c}\else \v{c}\fi{}i\ifmmode \check{c}\else
  \v{c}\fi{}evi\ifmmode~\acute{c}\else \'{c}\fi{}}}, \bibinfo {author}
  {\bibfnamefont {N.}~\bibnamefont {Wentzell}}, \bibinfo {author}
  {\bibfnamefont {M.}~\bibnamefont {Ferrero}}, \ and\ \bibinfo {author}
  {\bibfnamefont {O.}~\bibnamefont {Parcollet}},\ }\href {\doibase
  10.1103/PhysRevB.97.125141} {\bibfield  {journal} {\bibinfo  {journal} {Phys.
  Rev. B}\ }\textbf {\bibinfo {volume} {97}},\ \bibinfo {pages} {125141}
  (\bibinfo {year} {2018})}\BibitemShut {NoStop}%
\bibitem [{\citenamefont {Dahlen}\ and\ \citenamefont {van
  Leeuwen}(2005)}]{Dahlen05}%
  \BibitemOpen
  \bibfield  {author} {\bibinfo {author} {\bibfnamefont {N.~E.}\ \bibnamefont
  {Dahlen}}\ and\ \bibinfo {author} {\bibfnamefont {R.}~\bibnamefont {van
  Leeuwen}},\ }\href {\doibase http://dx.doi.org/10.1063/1.1884965} {\bibfield
  {journal} {\bibinfo  {journal} {The Journal of Chemical Physics}\ }\textbf
  {\bibinfo {volume} {122}},\ \bibinfo {eid} {164102} (\bibinfo {year}
  {2005}),\ http://dx.doi.org/10.1063/1.1884965}\BibitemShut {NoStop}%
\bibitem [{\citenamefont {Phillips}\ and\ \citenamefont {Zgid}(2014)}]{Zgid14}%
  \BibitemOpen
  \bibfield  {author} {\bibinfo {author} {\bibfnamefont {J.~J.}\ \bibnamefont
  {Phillips}}\ and\ \bibinfo {author} {\bibfnamefont {D.}~\bibnamefont
  {Zgid}},\ }\href {\doibase http://dx.doi.org/10.1063/1.4884951} {\bibfield
  {journal} {\bibinfo  {journal} {The Journal of Chemical Physics}\ }\textbf
  {\bibinfo {volume} {140}},\ \bibinfo {eid} {241101} (\bibinfo {year}
  {2014}),\ http://dx.doi.org/10.1063/1.4884951}\BibitemShut {NoStop}%
\bibitem [{\citenamefont {Rusakov}\ and\ \citenamefont
  {Zgid}(2016)}]{Rusakov16}%
  \BibitemOpen
  \bibfield  {author} {\bibinfo {author} {\bibfnamefont {A.~A.}\ \bibnamefont
  {Rusakov}}\ and\ \bibinfo {author} {\bibfnamefont {D.}~\bibnamefont {Zgid}},\
  }\href {\doibase http://dx.doi.org/10.1063/1.4940900} {\bibfield  {journal}
  {\bibinfo  {journal} {The Journal of Chemical Physics}\ }\textbf {\bibinfo
  {volume} {144}},\ \bibinfo {eid} {054106} (\bibinfo {year} {2016}),\
  http://dx.doi.org/10.1063/1.4940900}\BibitemShut {NoStop}%
\bibitem [{\citenamefont {Phillips}\ \emph {et~al.}(2015)\citenamefont
  {Phillips}, \citenamefont {Kananenka},\ and\ \citenamefont
  {Zgid}}]{Phillips15}%
  \BibitemOpen
  \bibfield  {author} {\bibinfo {author} {\bibfnamefont {J.~J.}\ \bibnamefont
  {Phillips}}, \bibinfo {author} {\bibfnamefont {A.~A.}\ \bibnamefont
  {Kananenka}}, \ and\ \bibinfo {author} {\bibfnamefont {D.}~\bibnamefont
  {Zgid}},\ }\href {\doibase http://dx.doi.org/10.1063/1.4921259} {\bibfield
  {journal} {\bibinfo  {journal} {The Journal of Chemical Physics}\ }\textbf
  {\bibinfo {volume} {142}},\ \bibinfo {eid} {194108} (\bibinfo {year}
  {2015}),\ http://dx.doi.org/10.1063/1.4921259}\BibitemShut {NoStop}%
\bibitem [{\citenamefont {Kananenka}\ \emph
  {et~al.}(2016{\natexlab{a}})\citenamefont {Kananenka}, \citenamefont
  {Phillips},\ and\ \citenamefont {Zgid}}]{Kananenka15}%
  \BibitemOpen
  \bibfield  {author} {\bibinfo {author} {\bibfnamefont {A.~A.}\ \bibnamefont
  {Kananenka}}, \bibinfo {author} {\bibfnamefont {J.~J.}\ \bibnamefont
  {Phillips}}, \ and\ \bibinfo {author} {\bibfnamefont {D.}~\bibnamefont
  {Zgid}},\ }\href {\doibase 10.1021/acs.jctc.5b00884} {\bibfield  {journal}
  {\bibinfo  {journal} {Journal of Chemical Theory and Computation}\ }\textbf
  {\bibinfo {volume} {12}},\ \bibinfo {pages} {564} (\bibinfo {year}
  {2016}{\natexlab{a}})},\ \Eprint
  {http://arxiv.org/abs/http://dx.doi.org/10.1021/acs.jctc.5b00884}
  {http://dx.doi.org/10.1021/acs.jctc.5b00884} \BibitemShut {NoStop}%
\bibitem [{\citenamefont {Kananenka}\ \emph
  {et~al.}(2016{\natexlab{b}})\citenamefont {Kananenka}, \citenamefont
  {Welden}, \citenamefont {Lan}, \citenamefont {Gull},\ and\ \citenamefont
  {Zgid}}]{Kananenka16}%
  \BibitemOpen
  \bibfield  {author} {\bibinfo {author} {\bibfnamefont {A.~A.}\ \bibnamefont
  {Kananenka}}, \bibinfo {author} {\bibfnamefont {A.~R.}\ \bibnamefont
  {Welden}}, \bibinfo {author} {\bibfnamefont {T.~N.}\ \bibnamefont {Lan}},
  \bibinfo {author} {\bibfnamefont {E.}~\bibnamefont {Gull}}, \ and\ \bibinfo
  {author} {\bibfnamefont {D.}~\bibnamefont {Zgid}},\ }\href {\doibase
  10.1021/acs.jctc.6b00178} {\bibfield  {journal} {\bibinfo  {journal} {Journal
  of Chemical Theory and Computation}\ }\textbf {\bibinfo {volume} {12}},\
  \bibinfo {pages} {2250} (\bibinfo {year} {2016}{\natexlab{b}})},\ \Eprint
  {http://arxiv.org/abs/http://dx.doi.org/10.1021/acs.jctc.6b00178}
  {http://dx.doi.org/10.1021/acs.jctc.6b00178} \BibitemShut {NoStop}%
\bibitem [{\citenamefont {Welden}\ \emph {et~al.}(2016)\citenamefont {Welden},
  \citenamefont {Rusakov},\ and\ \citenamefont {Zgid}}]{Welden16}%
  \BibitemOpen
  \bibfield  {author} {\bibinfo {author} {\bibfnamefont {A.~R.}\ \bibnamefont
  {Welden}}, \bibinfo {author} {\bibfnamefont {A.~A.}\ \bibnamefont {Rusakov}},
  \ and\ \bibinfo {author} {\bibfnamefont {D.}~\bibnamefont {Zgid}},\ }\href
  {\doibase 10.1063/1.4967449} {\bibfield  {journal} {\bibinfo  {journal} {The
  Journal of Chemical Physics}\ }\textbf {\bibinfo {volume} {145}},\ \bibinfo
  {pages} {204106} (\bibinfo {year} {2016})}\BibitemShut {NoStop}%
\bibitem [{\citenamefont {Kananenka}\ and\ \citenamefont
  {Zgid}(0)}]{kananenka_hybrif_gf2}%
  \BibitemOpen
  \bibfield  {author} {\bibinfo {author} {\bibfnamefont {A.~A.}\ \bibnamefont
  {Kananenka}}\ and\ \bibinfo {author} {\bibfnamefont {D.}~\bibnamefont
  {Zgid}},\ }\href {\doibase 10.1021/acs.jctc.7b00701} {\bibfield  {journal}
  {\bibinfo  {journal} {Journal of Chemical Theory and Computation}\ }\textbf
  {\bibinfo {volume} {0}},\ \bibinfo {pages} {null} (\bibinfo {year} {0})},\
  \bibinfo {note} {pMID: 28921986},\ \Eprint
  {http://arxiv.org/abs/http://dx.doi.org/10.1021/acs.jctc.7b00701}
  {http://dx.doi.org/10.1021/acs.jctc.7b00701} \BibitemShut {NoStop}%
\bibitem [{\citenamefont {Gull}\ \emph {et~al.}(2018)\citenamefont {Gull},
  \citenamefont {Iskakov}, \citenamefont {Krivenko}, \citenamefont {Rusakov},\
  and\ \citenamefont {Zgid}}]{Iskakov_Chebychev_2018}%
  \BibitemOpen
  \bibfield  {author} {\bibinfo {author} {\bibfnamefont {E.}~\bibnamefont
  {Gull}}, \bibinfo {author} {\bibfnamefont {S.}~\bibnamefont {Iskakov}},
  \bibinfo {author} {\bibfnamefont {I.}~\bibnamefont {Krivenko}}, \bibinfo
  {author} {\bibfnamefont {A.~A.}\ \bibnamefont {Rusakov}}, \ and\ \bibinfo
  {author} {\bibfnamefont {D.}~\bibnamefont {Zgid}},\ }\href {\doibase
  10.1103/PhysRevB.98.075127} {\bibfield  {journal} {\bibinfo  {journal} {Phys.
  Rev. B}\ }\textbf {\bibinfo {volume} {98}},\ \bibinfo {pages} {075127}
  (\bibinfo {year} {2018})}\BibitemShut {NoStop}%
\bibitem [{\citenamefont {Hedin}(1965)}]{Hedin65}%
  \BibitemOpen
  \bibfield  {author} {\bibinfo {author} {\bibfnamefont {L.}~\bibnamefont
  {Hedin}},\ }\href {\doibase 10.1103/PhysRev.139.A796} {\bibfield  {journal}
  {\bibinfo  {journal} {Phys. Rev.}\ }\textbf {\bibinfo {volume} {139}},\
  \bibinfo {pages} {A796} (\bibinfo {year} {1965})}\BibitemShut {NoStop}%
\bibitem [{\citenamefont {Gull}\ \emph {et~al.}(2011)\citenamefont {Gull},
  \citenamefont {Millis}, \citenamefont {Lichtenstein}, \citenamefont
  {Rubtsov}, \citenamefont {Troyer},\ and\ \citenamefont {Werner}}]{Gull11}%
  \BibitemOpen
  \bibfield  {author} {\bibinfo {author} {\bibfnamefont {E.}~\bibnamefont
  {Gull}}, \bibinfo {author} {\bibfnamefont {A.~J.}\ \bibnamefont {Millis}},
  \bibinfo {author} {\bibfnamefont {A.~I.}\ \bibnamefont {Lichtenstein}},
  \bibinfo {author} {\bibfnamefont {A.~N.}\ \bibnamefont {Rubtsov}}, \bibinfo
  {author} {\bibfnamefont {M.}~\bibnamefont {Troyer}}, \ and\ \bibinfo {author}
  {\bibfnamefont {P.}~\bibnamefont {Werner}},\ }\href {\doibase
  10.1103/RevModPhys.83.349} {\bibfield  {journal} {\bibinfo  {journal} {Rev.
  Mod. Phys.}\ }\textbf {\bibinfo {volume} {83}},\ \bibinfo {pages} {349}
  (\bibinfo {year} {2011})}\BibitemShut {NoStop}%
\bibitem [{\citenamefont {Zgid}\ and\ \citenamefont {Chan}(2011)}]{Zgid11}%
  \BibitemOpen
  \bibfield  {author} {\bibinfo {author} {\bibfnamefont {D.}~\bibnamefont
  {Zgid}}\ and\ \bibinfo {author} {\bibfnamefont {G.~K.-L.}\ \bibnamefont
  {Chan}},\ }\href {\doibase http://dx.doi.org/10.1063/1.3556707} {\bibfield
  {journal} {\bibinfo  {journal} {The Journal of Chemical Physics}\ }\textbf
  {\bibinfo {volume} {134}},\ \bibinfo {eid} {094115} (\bibinfo {year}
  {2011}),\ http://dx.doi.org/10.1063/1.3556707}\BibitemShut {NoStop}%
\bibitem [{\citenamefont {Zgid}\ \emph {et~al.}(2012)\citenamefont {Zgid},
  \citenamefont {Gull},\ and\ \citenamefont {Chan}}]{Zgid12}%
  \BibitemOpen
  \bibfield  {author} {\bibinfo {author} {\bibfnamefont {D.}~\bibnamefont
  {Zgid}}, \bibinfo {author} {\bibfnamefont {E.}~\bibnamefont {Gull}}, \ and\
  \bibinfo {author} {\bibfnamefont {G.~K.-L.}\ \bibnamefont {Chan}},\ }\href
  {\doibase 10.1103/PhysRevB.86.165128} {\bibfield  {journal} {\bibinfo
  {journal} {Phys. Rev. B}\ }\textbf {\bibinfo {volume} {86}},\ \bibinfo
  {pages} {165128} (\bibinfo {year} {2012})}\BibitemShut {NoStop}%
\bibitem [{\citenamefont {Blankenbecler}\ \emph {et~al.}(1981)\citenamefont
  {Blankenbecler}, \citenamefont {Scalapino},\ and\ \citenamefont
  {Sugar}}]{PhysRevD.24.2278}%
  \BibitemOpen
  \bibfield  {author} {\bibinfo {author} {\bibfnamefont {R.}~\bibnamefont
  {Blankenbecler}}, \bibinfo {author} {\bibfnamefont {D.~J.}\ \bibnamefont
  {Scalapino}}, \ and\ \bibinfo {author} {\bibfnamefont {R.~L.}\ \bibnamefont
  {Sugar}},\ }\href {\doibase 10.1103/PhysRevD.24.2278} {\bibfield  {journal}
  {\bibinfo  {journal} {Phys. Rev. D}\ }\textbf {\bibinfo {volume} {24}},\
  \bibinfo {pages} {2278} (\bibinfo {year} {1981})}\BibitemShut {NoStop}%
\bibitem [{\citenamefont {Zhang}\ \emph {et~al.}(1997)\citenamefont {Zhang},
  \citenamefont {Carlson},\ and\ \citenamefont
  {Gubernatis}}]{PhysRevB.55.7464}%
  \BibitemOpen
  \bibfield  {author} {\bibinfo {author} {\bibfnamefont {S.}~\bibnamefont
  {Zhang}}, \bibinfo {author} {\bibfnamefont {J.}~\bibnamefont {Carlson}}, \
  and\ \bibinfo {author} {\bibfnamefont {J.~E.}\ \bibnamefont {Gubernatis}},\
  }\href {\doibase 10.1103/PhysRevB.55.7464} {\bibfield  {journal} {\bibinfo
  {journal} {Phys. Rev. B}\ }\textbf {\bibinfo {volume} {55}},\ \bibinfo
  {pages} {7464} (\bibinfo {year} {1997})}\BibitemShut {NoStop}%
\bibitem [{\citenamefont {Zhang}\ and\ \citenamefont
  {Krakauer}(2003)}]{PhysRevLett.90.136401}%
  \BibitemOpen
  \bibfield  {author} {\bibinfo {author} {\bibfnamefont {S.}~\bibnamefont
  {Zhang}}\ and\ \bibinfo {author} {\bibfnamefont {H.}~\bibnamefont
  {Krakauer}},\ }\href {\doibase 10.1103/PhysRevLett.90.136401} {\bibfield
  {journal} {\bibinfo  {journal} {Phys. Rev. Lett.}\ }\textbf {\bibinfo
  {volume} {90}},\ \bibinfo {pages} {136401} (\bibinfo {year}
  {2003})}\BibitemShut {NoStop}%
\bibitem [{\citenamefont {Motta}\ and\ \citenamefont
  {Zhang}(2017)}]{doi:10.1021/acs.jctc.7b00730}%
  \BibitemOpen
  \bibfield  {author} {\bibinfo {author} {\bibfnamefont {M.}~\bibnamefont
  {Motta}}\ and\ \bibinfo {author} {\bibfnamefont {S.}~\bibnamefont {Zhang}},\
  }\href {\doibase 10.1021/acs.jctc.7b00730} {\bibfield  {journal} {\bibinfo
  {journal} {Journal of Chemical Theory and Computation}\ }\textbf {\bibinfo
  {volume} {13}},\ \bibinfo {pages} {5367} (\bibinfo {year} {2017})},\ \bibinfo
  {note} {pMID: 29053270},\ \Eprint
  {http://arxiv.org/abs/https://doi.org/10.1021/acs.jctc.7b00730}
  {https://doi.org/10.1021/acs.jctc.7b00730} \BibitemShut {NoStop}%
\bibitem [{\citenamefont {Marzari}\ \emph {et~al.}(2012)\citenamefont
  {Marzari}, \citenamefont {Mostofi}, \citenamefont {Yates}, \citenamefont
  {Souza},\ and\ \citenamefont {Vanderbilt}}]{RevModPhys.84.1419}%
  \BibitemOpen
  \bibfield  {author} {\bibinfo {author} {\bibfnamefont {N.}~\bibnamefont
  {Marzari}}, \bibinfo {author} {\bibfnamefont {A.~A.}\ \bibnamefont
  {Mostofi}}, \bibinfo {author} {\bibfnamefont {J.~R.}\ \bibnamefont {Yates}},
  \bibinfo {author} {\bibfnamefont {I.}~\bibnamefont {Souza}}, \ and\ \bibinfo
  {author} {\bibfnamefont {D.}~\bibnamefont {Vanderbilt}},\ }\href {\doibase
  10.1103/RevModPhys.84.1419} {\bibfield  {journal} {\bibinfo  {journal} {Rev.
  Mod. Phys.}\ }\textbf {\bibinfo {volume} {84}},\ \bibinfo {pages} {1419}
  (\bibinfo {year} {2012})}\BibitemShut {NoStop}%
\bibitem [{\citenamefont {Koch}\ and\ \citenamefont
  {Goedecker}(2001)}]{Koch_Goedecker}%
  \BibitemOpen
  \bibfield  {author} {\bibinfo {author} {\bibfnamefont {E.}~\bibnamefont
  {Koch}}\ and\ \bibinfo {author} {\bibfnamefont {S.}~\bibnamefont
  {Goedecker}},\ }\href {\doibase
  https://doi.org/10.1016/S0038-1098(01)00192-2} {\bibfield  {journal}
  {\bibinfo  {journal} {Solid State Communications}\ }\textbf {\bibinfo
  {volume} {119}},\ \bibinfo {pages} {105 } (\bibinfo {year}
  {2001})}\BibitemShut {NoStop}%
\bibitem [{\citenamefont {Kudinov}(1999)}]{Kudinov}%
  \BibitemOpen
  \bibfield  {author} {\bibinfo {author} {\bibfnamefont {E.}~\bibnamefont
  {Kudinov}},\ }\href@noop {} {\bibfield  {journal} {\bibinfo  {journal} {Fiz.
  Tverd. Tela (St. Petersburg) (Physics of the Solid State )}\ }\textbf
  {\bibinfo {volume} {41}},\ \bibinfo {pages} {1582} (\bibinfo {year}
  {1999})}\BibitemShut {NoStop}%
\bibitem [{\citenamefont {Goedecker}(1998)}]{Goedecker}%
  \BibitemOpen
  \bibfield  {author} {\bibinfo {author} {\bibfnamefont {S.}~\bibnamefont
  {Goedecker}},\ }\href {\doibase 10.1103/PhysRevB.58.3501} {\bibfield
  {journal} {\bibinfo  {journal} {Phys. Rev. B}\ }\textbf {\bibinfo {volume}
  {58}},\ \bibinfo {pages} {3501} (\bibinfo {year} {1998})}\BibitemShut
  {NoStop}%
\bibitem [{\citenamefont {Rusakov}\ \emph {et~al.}(2014)\citenamefont
  {Rusakov}, \citenamefont {Phillips},\ and\ \citenamefont {Zgid}}]{Rusakov14}%
  \BibitemOpen
  \bibfield  {author} {\bibinfo {author} {\bibfnamefont {A.~A.}\ \bibnamefont
  {Rusakov}}, \bibinfo {author} {\bibfnamefont {J.~J.}\ \bibnamefont
  {Phillips}}, \ and\ \bibinfo {author} {\bibfnamefont {D.}~\bibnamefont
  {Zgid}},\ }\href {\doibase http://dx.doi.org/10.1063/1.4901432} {\bibfield
  {journal} {\bibinfo  {journal} {The Journal of Chemical Physics}\ }\textbf
  {\bibinfo {volume} {141}},\ \bibinfo {eid} {194105} (\bibinfo {year}
  {2014}),\ http://dx.doi.org/10.1063/1.4901432}\BibitemShut {NoStop}%
\bibitem [{\citenamefont {Hachmann}\ \emph {et~al.}(2006)\citenamefont
  {Hachmann}, \citenamefont {Cardoen},\ and\ \citenamefont
  {Chan}}]{Hachmann:jcp2006-h50dmrg}%
  \BibitemOpen
  \bibfield  {author} {\bibinfo {author} {\bibfnamefont {J.}~\bibnamefont
  {Hachmann}}, \bibinfo {author} {\bibfnamefont {W.}~\bibnamefont {Cardoen}}, \
  and\ \bibinfo {author} {\bibfnamefont {G.~K.-L.}\ \bibnamefont {Chan}},\
  }\href@noop {} {\bibfield  {journal} {\bibinfo  {journal} {J. Chem. Phys.}\
  }\textbf {\bibinfo {volume} {125}},\ \bibinfo {pages} {144101} (\bibinfo
  {year} {2006})}\BibitemShut {NoStop}%
\bibitem [{\citenamefont {Frisch}\ \emph {et~al.}()\citenamefont {Frisch},
  \citenamefont {Trucks}, \citenamefont {Schlegel}, \citenamefont {Scuseria},
  \citenamefont {Robb}, \citenamefont {Cheeseman}, \citenamefont {Scalmani},
  \citenamefont {Barone}, \citenamefont {Mennucci}, \citenamefont {Petersson},
  \citenamefont {Nakatsuji}, \citenamefont {Caricato}, \citenamefont {Li},
  \citenamefont {Hratchian}, \citenamefont {Izmaylov}, \citenamefont {Bloino},
  \citenamefont {Zheng}, \citenamefont {Sonnenberg}, \citenamefont {Hada},
  \citenamefont {Ehara}, \citenamefont {Toyota}, \citenamefont {Fukuda},
  \citenamefont {Hasegawa}, \citenamefont {Ishida}, \citenamefont {Nakajima},
  \citenamefont {Honda}, \citenamefont {Kitao}, \citenamefont {Nakai},
  \citenamefont {Vreven}, \citenamefont {Montgomery}, \citenamefont {Peralta},
  \citenamefont {Ogliaro}, \citenamefont {Bearpark}, \citenamefont {Heyd},
  \citenamefont {Brothers}, \citenamefont {Kudin}, \citenamefont {Staroverov},
  \citenamefont {Kobayashi}, \citenamefont {Normand}, \citenamefont
  {Raghavachari}, \citenamefont {Rendell}, \citenamefont {Burant},
  \citenamefont {Iyengar}, \citenamefont {Tomasi}, \citenamefont {Cossi},
  \citenamefont {Rega}, \citenamefont {Millam}, \citenamefont {Klene},
  \citenamefont {Knox}, \citenamefont {Cross}, \citenamefont {Bakken},
  \citenamefont {Adamo}, \citenamefont {Jaramillo}, \citenamefont {Gomperts},
  \citenamefont {Stratmann}, \citenamefont {Yazyev}, \citenamefont {Austin},
  \citenamefont {Cammi}, \citenamefont {Pomelli}, \citenamefont {Ochterski},
  \citenamefont {Martin}, \citenamefont {Morokuma}, \citenamefont {Zakrzewski},
  \citenamefont {Voth}, \citenamefont {Salvador}, \citenamefont {Dannenberg},
  \citenamefont {Dapprich}, \citenamefont {Daniels}, \citenamefont {Farkas},
  \citenamefont {Foresman}, \citenamefont {Ortiz}, \citenamefont {Cioslowski},\
  and\ \citenamefont {Fox}}]{g09}%
  \BibitemOpen
  \bibfield  {author} {\bibinfo {author} {\bibfnamefont {M.~J.}\ \bibnamefont
  {Frisch}}, \bibinfo {author} {\bibfnamefont {G.~W.}\ \bibnamefont {Trucks}},
  \bibinfo {author} {\bibfnamefont {H.~B.}\ \bibnamefont {Schlegel}}, \bibinfo
  {author} {\bibfnamefont {G.~E.}\ \bibnamefont {Scuseria}}, \bibinfo {author}
  {\bibfnamefont {M.~A.}\ \bibnamefont {Robb}}, \bibinfo {author}
  {\bibfnamefont {J.~R.}\ \bibnamefont {Cheeseman}}, \bibinfo {author}
  {\bibfnamefont {G.}~\bibnamefont {Scalmani}}, \bibinfo {author}
  {\bibfnamefont {V.}~\bibnamefont {Barone}}, \bibinfo {author} {\bibfnamefont
  {B.}~\bibnamefont {Mennucci}}, \bibinfo {author} {\bibfnamefont {G.~A.}\
  \bibnamefont {Petersson}}, \bibinfo {author} {\bibfnamefont {H.}~\bibnamefont
  {Nakatsuji}}, \bibinfo {author} {\bibfnamefont {M.}~\bibnamefont {Caricato}},
  \bibinfo {author} {\bibfnamefont {X.}~\bibnamefont {Li}}, \bibinfo {author}
  {\bibfnamefont {H.~P.}\ \bibnamefont {Hratchian}}, \bibinfo {author}
  {\bibfnamefont {A.~F.}\ \bibnamefont {Izmaylov}}, \bibinfo {author}
  {\bibfnamefont {J.}~\bibnamefont {Bloino}}, \bibinfo {author} {\bibfnamefont
  {G.}~\bibnamefont {Zheng}}, \bibinfo {author} {\bibfnamefont {J.~L.}\
  \bibnamefont {Sonnenberg}}, \bibinfo {author} {\bibfnamefont
  {M.}~\bibnamefont {Hada}}, \bibinfo {author} {\bibfnamefont {M.}~\bibnamefont
  {Ehara}}, \bibinfo {author} {\bibfnamefont {K.}~\bibnamefont {Toyota}},
  \bibinfo {author} {\bibfnamefont {R.}~\bibnamefont {Fukuda}}, \bibinfo
  {author} {\bibfnamefont {J.}~\bibnamefont {Hasegawa}}, \bibinfo {author}
  {\bibfnamefont {M.}~\bibnamefont {Ishida}}, \bibinfo {author} {\bibfnamefont
  {T.}~\bibnamefont {Nakajima}}, \bibinfo {author} {\bibfnamefont
  {Y.}~\bibnamefont {Honda}}, \bibinfo {author} {\bibfnamefont
  {O.}~\bibnamefont {Kitao}}, \bibinfo {author} {\bibfnamefont
  {H.}~\bibnamefont {Nakai}}, \bibinfo {author} {\bibfnamefont
  {T.}~\bibnamefont {Vreven}}, \bibinfo {author} {\bibfnamefont {J.~A.}\
  \bibnamefont {Montgomery}, \bibfnamefont {{Jr.}}}, \bibinfo {author}
  {\bibfnamefont {J.~E.}\ \bibnamefont {Peralta}}, \bibinfo {author}
  {\bibfnamefont {F.}~\bibnamefont {Ogliaro}}, \bibinfo {author} {\bibfnamefont
  {M.}~\bibnamefont {Bearpark}}, \bibinfo {author} {\bibfnamefont {J.~J.}\
  \bibnamefont {Heyd}}, \bibinfo {author} {\bibfnamefont {E.}~\bibnamefont
  {Brothers}}, \bibinfo {author} {\bibfnamefont {K.~N.}\ \bibnamefont {Kudin}},
  \bibinfo {author} {\bibfnamefont {V.~N.}\ \bibnamefont {Staroverov}},
  \bibinfo {author} {\bibfnamefont {R.}~\bibnamefont {Kobayashi}}, \bibinfo
  {author} {\bibfnamefont {J.}~\bibnamefont {Normand}}, \bibinfo {author}
  {\bibfnamefont {K.}~\bibnamefont {Raghavachari}}, \bibinfo {author}
  {\bibfnamefont {A.}~\bibnamefont {Rendell}}, \bibinfo {author} {\bibfnamefont
  {J.~C.}\ \bibnamefont {Burant}}, \bibinfo {author} {\bibfnamefont {S.~S.}\
  \bibnamefont {Iyengar}}, \bibinfo {author} {\bibfnamefont {J.}~\bibnamefont
  {Tomasi}}, \bibinfo {author} {\bibfnamefont {M.}~\bibnamefont {Cossi}},
  \bibinfo {author} {\bibfnamefont {N.}~\bibnamefont {Rega}}, \bibinfo {author}
  {\bibfnamefont {J.~M.}\ \bibnamefont {Millam}}, \bibinfo {author}
  {\bibfnamefont {M.}~\bibnamefont {Klene}}, \bibinfo {author} {\bibfnamefont
  {J.~E.}\ \bibnamefont {Knox}}, \bibinfo {author} {\bibfnamefont {J.~B.}\
  \bibnamefont {Cross}}, \bibinfo {author} {\bibfnamefont {V.}~\bibnamefont
  {Bakken}}, \bibinfo {author} {\bibfnamefont {C.}~\bibnamefont {Adamo}},
  \bibinfo {author} {\bibfnamefont {J.}~\bibnamefont {Jaramillo}}, \bibinfo
  {author} {\bibfnamefont {R.}~\bibnamefont {Gomperts}}, \bibinfo {author}
  {\bibfnamefont {R.~E.}\ \bibnamefont {Stratmann}}, \bibinfo {author}
  {\bibfnamefont {O.}~\bibnamefont {Yazyev}}, \bibinfo {author} {\bibfnamefont
  {A.~J.}\ \bibnamefont {Austin}}, \bibinfo {author} {\bibfnamefont
  {R.}~\bibnamefont {Cammi}}, \bibinfo {author} {\bibfnamefont
  {C.}~\bibnamefont {Pomelli}}, \bibinfo {author} {\bibfnamefont {J.~W.}\
  \bibnamefont {Ochterski}}, \bibinfo {author} {\bibfnamefont {R.~L.}\
  \bibnamefont {Martin}}, \bibinfo {author} {\bibfnamefont {K.}~\bibnamefont
  {Morokuma}}, \bibinfo {author} {\bibfnamefont {V.~G.}\ \bibnamefont
  {Zakrzewski}}, \bibinfo {author} {\bibfnamefont {G.~A.}\ \bibnamefont
  {Voth}}, \bibinfo {author} {\bibfnamefont {P.}~\bibnamefont {Salvador}},
  \bibinfo {author} {\bibfnamefont {J.~J.}\ \bibnamefont {Dannenberg}},
  \bibinfo {author} {\bibfnamefont {S.}~\bibnamefont {Dapprich}}, \bibinfo
  {author} {\bibfnamefont {A.~D.}\ \bibnamefont {Daniels}}, \bibinfo {author}
  {\bibfnamefont {{\"O}.}~\bibnamefont {Farkas}}, \bibinfo {author}
  {\bibfnamefont {J.~B.}\ \bibnamefont {Foresman}}, \bibinfo {author}
  {\bibfnamefont {J.~V.}\ \bibnamefont {Ortiz}}, \bibinfo {author}
  {\bibfnamefont {J.}~\bibnamefont {Cioslowski}}, \ and\ \bibinfo {author}
  {\bibfnamefont {D.~J.}\ \bibnamefont {Fox}},\ }\href@noop {} {\enquote
  {\bibinfo {title} {{Gaussian09} {R}evision {A}.02},}\ }\bibinfo {note}
  {Gaussian Inc. Wallingford CT 2009}\BibitemShut {NoStop}%
\bibitem [{\citenamefont {Levy}\ \emph {et~al.}(2017)\citenamefont {Levy},
  \citenamefont {LeBlanc},\ and\ \citenamefont {Gull}}]{LEVY2017149}%
  \BibitemOpen
  \bibfield  {author} {\bibinfo {author} {\bibfnamefont {R.}~\bibnamefont
  {Levy}}, \bibinfo {author} {\bibfnamefont {J.}~\bibnamefont {LeBlanc}}, \
  and\ \bibinfo {author} {\bibfnamefont {E.}~\bibnamefont {Gull}},\ }\href
  {\doibase https://doi.org/10.1016/j.cpc.2017.01.018} {\bibfield  {journal}
  {\bibinfo  {journal} {Computer Physics Communications}\ }\textbf {\bibinfo
  {volume} {215}},\ \bibinfo {pages} {149 } (\bibinfo {year}
  {2017})}\BibitemShut {NoStop}%
\bibitem [{\citenamefont {Nilsson}\ \emph {et~al.}(2017)\citenamefont
  {Nilsson}, \citenamefont {Boehnke}, \citenamefont {Werner},\ and\
  \citenamefont {Aryasetiawan}}]{PhysRevMaterials.1.043803}%
  \BibitemOpen
  \bibfield  {author} {\bibinfo {author} {\bibfnamefont {F.}~\bibnamefont
  {Nilsson}}, \bibinfo {author} {\bibfnamefont {L.}~\bibnamefont {Boehnke}},
  \bibinfo {author} {\bibfnamefont {P.}~\bibnamefont {Werner}}, \ and\ \bibinfo
  {author} {\bibfnamefont {F.}~\bibnamefont {Aryasetiawan}},\ }\href {\doibase
  10.1103/PhysRevMaterials.1.043803} {\bibfield  {journal} {\bibinfo  {journal}
  {Phys. Rev. Materials}\ }\textbf {\bibinfo {volume} {1}},\ \bibinfo {pages}
  {043803} (\bibinfo {year} {2017})}\BibitemShut {NoStop}%
\bibitem [{\citenamefont {Ayral}\ \emph {et~al.}(2017)\citenamefont {Ayral},
  \citenamefont {Biermann}, \citenamefont {Werner},\ and\ \citenamefont
  {Boehnke}}]{PhysRevB.95.245130}%
  \BibitemOpen
  \bibfield  {author} {\bibinfo {author} {\bibfnamefont {T.}~\bibnamefont
  {Ayral}}, \bibinfo {author} {\bibfnamefont {S.}~\bibnamefont {Biermann}},
  \bibinfo {author} {\bibfnamefont {P.}~\bibnamefont {Werner}}, \ and\ \bibinfo
  {author} {\bibfnamefont {L.}~\bibnamefont {Boehnke}},\ }\href {\doibase
  10.1103/PhysRevB.95.245130} {\bibfield  {journal} {\bibinfo  {journal} {Phys.
  Rev. B}\ }\textbf {\bibinfo {volume} {95}},\ \bibinfo {pages} {245130}
  (\bibinfo {year} {2017})}\BibitemShut {NoStop}%
\bibitem [{\citenamefont {Boehnke}\ \emph {et~al.}(2016)\citenamefont
  {Boehnke}, \citenamefont {Nilsson}, \citenamefont {Aryasetiawan},\ and\
  \citenamefont {Werner}}]{PhysRevB.94.201106}%
  \BibitemOpen
  \bibfield  {author} {\bibinfo {author} {\bibfnamefont {L.}~\bibnamefont
  {Boehnke}}, \bibinfo {author} {\bibfnamefont {F.}~\bibnamefont {Nilsson}},
  \bibinfo {author} {\bibfnamefont {F.}~\bibnamefont {Aryasetiawan}}, \ and\
  \bibinfo {author} {\bibfnamefont {P.}~\bibnamefont {Werner}},\ }\href
  {\doibase 10.1103/PhysRevB.94.201106} {\bibfield  {journal} {\bibinfo
  {journal} {Phys. Rev. B}\ }\textbf {\bibinfo {volume} {94}},\ \bibinfo
  {pages} {201106} (\bibinfo {year} {2016})}\BibitemShut {NoStop}%
\end{thebibliography}
%

\end{document}